\documentclass[aps,prc,amsfonts,superscriptaddress,reprint,showpacs,showkeys]{revtex4-2}  

\usepackage{graphicx}  
\usepackage{amssymb}   
\usepackage{amsmath}
\usepackage{hyperref}
\usepackage{verbatim}
\usepackage{multirow}
\usepackage{siunitx}
\usepackage[modulo]{lineno}

\begin{document}

\title{Mass measurements of As, Se and Br nuclei and their implication on the proton-neutron interaction strength towards the N=Z line}



\author{I. Mardor}
\email[Corresponding author: ]{mardor@tauex.tau.ac.il}
\affiliation{Tel Aviv University, 6997801 Tel Aviv, Israel}
\affiliation{Soreq Nuclear Research Center, 81800 Yavne, Israel}
\author{S. Ayet San Andr\'{e}s}
\affiliation{GSI Helmholtzzentrum f{\"u}r Schwerionenforschung GmbH, 64291 Darmstadt, Germany}
\author{T. Dickel}
\affiliation{GSI Helmholtzzentrum f{\"u}r Schwerionenforschung GmbH, 64291 Darmstadt, Germany}
\affiliation{II.~Physikalisches Institut, Justus-Liebig-Universit{\"a}t Gie{\ss}en, 35392 Gie{\ss}en, Germany}
\author{D. Amanbayev}
\affiliation{II.~Physikalisches Institut, Justus-Liebig-Universit{\"a}t Gie{\ss}en, 35392 Gie{\ss}en, Germany}
\author{S. Beck}
\affiliation{GSI Helmholtzzentrum f{\"u}r Schwerionenforschung GmbH, 64291 Darmstadt, Germany}
\affiliation{II.~Physikalisches Institut, Justus-Liebig-Universit{\"a}t Gie{\ss}en, 35392 Gie{\ss}en, Germany}
\author{J. Bergmann}
\affiliation{II.~Physikalisches Institut, Justus-Liebig-Universit{\"a}t Gie{\ss}en, 35392 Gie{\ss}en, Germany}
\author{H. Geissel}
\affiliation{GSI Helmholtzzentrum f{\"u}r Schwerionenforschung GmbH, 64291 Darmstadt, Germany}
\affiliation{II.~Physikalisches Institut, Justus-Liebig-Universit{\"a}t Gie{\ss}en, 35392 Gie{\ss}en, Germany}
\author{L. Gr\"of}
\affiliation{II.~Physikalisches Institut, Justus-Liebig-Universit{\"a}t Gie{\ss}en, 35392 Gie{\ss}en, Germany}
\author{E. Haettner}
\affiliation{GSI Helmholtzzentrum f{\"u}r Schwerionenforschung GmbH, 64291 Darmstadt, Germany}
\author{C. Hornung}
\affiliation{II.~Physikalisches Institut, Justus-Liebig-Universit{\"a}t Gie{\ss}en, 35392 Gie{\ss}en, Germany}
\author{N. Kalantar-Nayestanaki}
\affiliation{Nuclear Energy Group, ESRIG, University of Groningen, 9747 AA  Groningen, The Netherlands}
\author{G. Kripko-Koncz}
\affiliation{II.~Physikalisches Institut, Justus-Liebig-Universit{\"a}t Gie{\ss}en, 35392 Gie{\ss}en, Germany}
\author{I. Miskun}
\affiliation{II.~Physikalisches Institut, Justus-Liebig-Universit{\"a}t Gie{\ss}en, 35392 Gie{\ss}en, Germany}
\author{A. Mollaebrahimi}
\affiliation{Nuclear Energy Group, ESRIG, University of Groningen, 9747 AA  Groningen, The Netherlands}
\affiliation{II.~Physikalisches Institut, Justus-Liebig-Universit{\"a}t Gie{\ss}en, 35392 Gie{\ss}en, Germany}
\author{W. R.\ Pla\ss}
\affiliation{GSI Helmholtzzentrum f{\"u}r Schwerionenforschung GmbH, 64291 Darmstadt, Germany}
\affiliation{II.~Physikalisches Institut, Justus-Liebig-Universit{\"a}t Gie{\ss}en, 35392 Gie{\ss}en, Germany}
\author{C. Scheidenberger}
\affiliation{GSI Helmholtzzentrum f{\"u}r Schwerionenforschung GmbH, 64291 Darmstadt, Germany}
\affiliation{II.~Physikalisches Institut, Justus-Liebig-Universit{\"a}t Gie{\ss}en, 35392 Gie{\ss}en, Germany}
\author{H. Weick}
\affiliation{GSI Helmholtzzentrum f{\"u}r Schwerionenforschung GmbH, 64291 Darmstadt, Germany}
\author{Soumya Bagchi} 
\email[Present address:]{ Indian Institute of Technology (Indian School of Mines), Dhanbad, Jharkhand 826004, India}
\affiliation{Saint Mary's University, NS B3H 3C3 Halifax, Canada}
\affiliation{GSI Helmholtzzentrum f{\"u}r Schwerionenforschung GmbH, 64291 Darmstadt, Germany}
\affiliation{II.~Physikalisches Institut, Justus-Liebig-Universit{\"a}t Gie{\ss}en, 35392 Gie{\ss}en, Germany}
\author{D. L. Balabanski}
\affiliation{Extreme Light Infrastructure-Nuclear Physics (ELI-NP), Horia Hulubei National Institute for R\&D in Physics and Nuclear Engineering, Str. Reactorului 30, 077125 Bucharest-M$\breve{a}$gurele, Romania}
\author{A. A. Bezbakh}
\affiliation{Flerov Laboratory of Nuclear Reactions, JINR, 141980 Dubna, Russia}
\affiliation{Institute of Physics, Silesian University in Opava, 74601 Opava, Czech Republic}
\author{Z. Brencic}
\affiliation{Jozef Stefan Institute, SI-1000 Ljubljana, Slovenia}
\author{O. Charviakova}
\affiliation{National Centre for Nuclear Research, Ho\.{z}a 69, 00-681 Warszawa, Poland}
\author{V. Chudoba}
\affiliation{Flerov Laboratory of Nuclear Reactions, JINR, 141980 Dubna, Russia}
\affiliation{Institute of Physics, Silesian University in Opava, 74601 Opava, Czech Republic}
\author{Paul Constantin} 
\affiliation{Extreme Light Infrastructure-Nuclear Physics (ELI-NP), Horia Hulubei National Institute for R\&D in Physics and Nuclear Engineering, Str. Reactorului 30, 077125 Bucharest-M$\breve{a}$gurele, Romania}
\author{M. Dehghan}
\affiliation{GSI Helmholtzzentrum f{\"u}r Schwerionenforschung GmbH, 64291 Darmstadt, Germany}
\author{A. S. Fomichev}
\affiliation{Flerov Laboratory of Nuclear Reactions, JINR, 141980 Dubna, Russia}
\author{L. V. Grigorenko}
\affiliation{Flerov Laboratory of Nuclear Reactions, JINR, 141980 Dubna, Russia}
\affiliation{National Research Centre “Kurchatov Institute”, 123182 Moscow, Russia}
\affiliation{National Research Nuclear University “MEPhI”, 115409 Moscow, Russia}
\author{O. Hall}
\affiliation{University of Edinburgh, EH8 9AB Edinburgh, United Kingdom}
\author{M. N. Harakeh}
\affiliation{Nuclear Energy Group, ESRIG, University of Groningen, 9747 AA  Groningen, The Netherlands}
\author{J.-P. Hucka}
\affiliation{GSI Helmholtzzentrum f{\"u}r Schwerionenforschung GmbH, 64291 Darmstadt, Germany}
\affiliation{Technische Universit\"{a}t Darmstadt, D-64289 Darmstadt, Germany}
\author{A. Kankainen}
\affiliation{University of Jyv\"askyl\"a, 40014 Jyv\"askyl\"a, Finland}
\affiliation{Helsinki Institute of Physics, 00014 Helsinki, Finland}
\author{O. Kiselev}
\affiliation{GSI Helmholtzzentrum f{\"u}r Schwerionenforschung GmbH, 64291 Darmstadt, Germany}
\author{R. Kn\"obel}
\affiliation{GSI Helmholtzzentrum f{\"u}r Schwerionenforschung GmbH, 64291 Darmstadt, Germany}
\author{D. A. Kostyleva}
\affiliation{GSI Helmholtzzentrum f{\"u}r Schwerionenforschung GmbH, 64291 Darmstadt, Germany}
\affiliation{II.~Physikalisches Institut, Justus-Liebig-Universit{\"a}t Gie{\ss}en, 35392 Gie{\ss}en, Germany}
\author{S. A. Krupko}
\affiliation{Flerov Laboratory of Nuclear Reactions, JINR, 141980 Dubna, Russia}
\affiliation{Institute of Physics, Silesian University in Opava, 74601 Opava, Czech Republic}
\author{N. Kurkova}
\affiliation{Flerov Laboratory of Nuclear Reactions, JINR, 141980 Dubna, Russia}
\author{N. Kuzminchuk}
\affiliation{GSI Helmholtzzentrum f{\"u}r Schwerionenforschung GmbH, 64291 Darmstadt, Germany}
\author{I. Mukha}
\affiliation{GSI Helmholtzzentrum f{\"u}r Schwerionenforschung GmbH, 64291 Darmstadt, Germany}
\author{I. A. Muzalevskii}
\affiliation{Flerov Laboratory of Nuclear Reactions, JINR, 141980 Dubna, Russia}
\affiliation{Institute of Physics, Silesian University in Opava, 74601 Opava, Czech Republic}
\author{D. Nichita}
\affiliation{Extreme Light Infrastructure-Nuclear Physics (ELI-NP), Horia Hulubei National Institute for R\&D in Physics and Nuclear Engineering, Str. Reactorului 30, 077125 Bucharest-M$\breve{a}$gurele, Romania}
\affiliation{Doctoral School in Engineering and Applications of Lasers and Accelerators, University Polytechnica of Bucharest, 060811 Bucharest, Romania}
\author{C. Nociforo}
\affiliation{GSI Helmholtzzentrum f{\"u}r Schwerionenforschung GmbH, 64291 Darmstadt, Germany}
\author{Z. Patyk}
\affiliation{National Centre for Nuclear Research, Ho\.{z}a 69, 00-681 Warszawa, Poland}
\author{M. Pf\"utzner}
\affiliation{Faculty of Physics, University of Warsaw, 02-093 Warszawa, Poland}
\author{S. Pietri}
\affiliation{GSI Helmholtzzentrum f{\"u}r Schwerionenforschung GmbH, 64291 Darmstadt, Germany}
\author{S. Purushothaman}
\affiliation{GSI Helmholtzzentrum f{\"u}r Schwerionenforschung GmbH, 64291 Darmstadt, Germany}
\author{M. P. Reiter}
\affiliation{University of Edinburgh, EH8 9AB Edinburgh, United Kingdom}
\author{H. Roesch}
\affiliation{GSI Helmholtzzentrum f{\"u}r Schwerionenforschung GmbH, 64291 Darmstadt, Germany}
\affiliation{Technische Universit\"{a}t Darmstadt, D-64289 Darmstadt, Germany}
\author{F. Schirru}
\affiliation{GSI Helmholtzzentrum f{\"u}r Schwerionenforschung GmbH, 64291 Darmstadt, Germany}
\author{P. G. Sharov}
\affiliation{Flerov Laboratory of Nuclear Reactions, JINR, 141980 Dubna, Russia}
\affiliation{Institute of Physics, Silesian University in Opava, 74601 Opava, Czech Republic}
\author{A. Sp\u{a}taru}
\affiliation{Extreme Light Infrastructure-Nuclear Physics (ELI-NP), Horia Hulubei National Institute for R\&D in Physics and Nuclear Engineering, Str. Reactorului 30, 077125 Bucharest-M$\breve{a}$gurele, Romania}
\affiliation{Doctoral School in Engineering and Applications of Lasers and Accelerators, University Polytechnica of Bucharest, 060811 Bucharest, Romania}
\author{G. Stanic}
\affiliation{Johannes Gutenberg-Universit\"{a}t Mainz, 55099 Mainz, Germany}
\author{A. State}
\affiliation{Extreme Light Infrastructure-Nuclear Physics (ELI-NP), Horia Hulubei National Institute for R\&D in Physics and Nuclear Engineering, Str. Reactorului 30, 077125 Bucharest-M$\breve{a}$gurele, Romania}
\affiliation{Doctoral School in Engineering and Applications of Lasers and Accelerators, University Polytechnica of Bucharest, 060811 Bucharest, Romania}
\author{Y. K. Tanaka}
\affiliation{High Energy Nuclear Physics Laboratory, RIKEN, 2-1 Hirosawa, Wako, 351-0198 Saitama, Japan}
\author{M. Vencelj}
\affiliation{Jozef Stefan Institute, SI-1000 Ljubljana, Slovenia}
\author{M. I.\ Yavor} 
\affiliation{Institute for Analytical Instrumentation, RAS, 190103 St. Petersburg, Russia}
\author{J. Zhao}
\affiliation{GSI Helmholtzzentrum f{\"u}r Schwerionenforschung GmbH, 64291 Darmstadt, Germany}

\date{\today}

\begin{abstract}
\vspace*{0.75cm}
Mass measurements of the nuclides $^{69}$As, $^{70,71}$Se and $^{71}$Br, produced via fragmentation of a $^{124}$Xe primary beam at the FRS at GSI, have been performed with the multiple-reflection time-of-flight mass spectrometer (MR-TOF-MS) of the FRS Ion Catcher with an unprecedented mass resolving power of almost 1,000,000. Such high resolving power is the only way to achieve accurate results and resolve overlapping peaks of short-lived exotic nuclei, whose total number of accumulated events is always limited.
For the nuclide $^{69}$As, this is the first direct mass measurement. A mass uncertainty of 22 keV was achieved with only 10 events. For the nuclide $^{70}$Se, a mass uncertainty of 2.6 keV was obtained, corresponding to a relative accuracy of $\delta m/m$ = 4.0$\times 10^{-8}$, with less than 500 events. The masses of the nuclides $^{71}$Se and $^{71}$Br have been measured with an uncertainty of 23 and 16 keV, respectively.
Our results for the nuclides $^{70,71}$Se and $^{71}$Br are in good agreement with the 2016 Atomic Mass Evaluation, and our result for the nuclide $^{69}$As resolves the discrepancy between the previous indirect measurements.
We measured also the mass of the molecule $^{14}$N$^{15}$N$^{40}$Ar (A=69) with a relative accuracy of $\delta m/m$ = 1.7$\times 10^{-8}$, the highest yet achieved with an MR-TOF-MS.
Our results show that the measured restrengthening of the proton-neutron interaction ($\delta$V$_{pn}$) for odd-odd nuclei along the N=Z line above Z=29 (recently extended to Z=37) is hardly evident at the N-Z=2 line, and not evident at the N-Z=4 line. Nevertheless, detailed structure of $\delta$V$_{pn}$ along the N-Z=2 and N-Z=4 lines, confirmed by our mass measurements, may provide a hint regarding the ongoing $\approx$500 keV discrepancy in the mass value of the nuclide $^{70}$Br, which prevents including it in the world average of ${Ft}$-value for superallowed 0$^+\rightarrow$ 0$^+$ $\beta$ decays. The reported work sets the stage for mass measurements with the FRS Ion Catcher of nuclei at and beyond the N=Z line in the same region of the nuclear chart, including the nuclide $^{70}$Br.
\end{abstract}


\maketitle


\section{Introduction}
\label{sec:intro}
Nuclear masses are key properties of atomic nuclei, as they are a measure of the nuclear binding energy, which reflects the details of nuclear structure and forces between the nucleons.
The extension of nuclear mass measurements towards rare, short-lived nuclei far from the valley of stability is important for widening nuclear structure models to all regions of the nuclear chart \cite{Dilling2018,zhang2016} and for astrophysical nucleosynthesis calculations \cite{Schatz2013}.

Nuclear masses of nuclei around the N=Z line provide crucial input for determining the path of the rapid proton-capture (rp-) process that takes place in Type-I X-ray bursts and steady-state nuclear burning in rapidly accreting neutron stars \cite{Schatz2017,Duy2020}.

Experimental masses of numerous odd-odd N=Z nuclei, which decay via super-allowed $0^+\rightarrow 0^+$ $\beta^+$-transitions, determine $Q_{EC}$ values for these transitions. These values, together with the half-lives and decay branching ratios, formulate the ${Ft}$-values that in turn test the unitarity of the CKM matrix, a fundamental principle of the Standard Model \cite{Hardy2015,Hardy2020}.

However, the ${Ft}$-value of $^{70}$Br that is derived from its precise half-life and decay branching ratio \cite{Morales2017} and only direct mass measurement (Penning trap \cite{Savory2009}) deviates by approximately 12$\sigma$ from the world average. Since the measurements of Ref. \cite{Morales2017} are accurate enough and not under dispute, the source of this discrepancy must lie in its mass value.
The mass measurement of Ref. \cite{Savory2009} differs by approximately 500 keV from that obtained via positron end-point energy \cite{Davids1980}, which gives a consistent ${Ft}$-value, but with a large uncertainty.

Actually, the Penning trap result was derived from measuring the mass of the $^{70m}$Br isomer ($t_{1/2}$ = 2.2 s) and subtracting its known excitation energy, because the cycle time employed in the measurement \cite{Savory2009} was much longer compared to the half-life of the nuclide $^{70}$Br ((78.42 $\pm$ 0.51) ms \cite{Morales2017}). Therefore, the Penning trap measurement of Ref.  \cite{Savory2009} is not included in the current ${Ft}$ value world average, and a new direct mass measurement of the nuclide $^{70}$Br is called for in Ref. \cite{Hardy2015,Hardy2020}.

In this article, we focus on insights that can be derived from accurate mass measurements on the interaction strength between the last proton and the last neutron of intermediate-mass nuclei towards the N=Z line. By 'last' we refer to the nucleons that occupy the highest energy orbitals in the nucleus.
This interaction strength, denoted as $\delta$V$_{pn}$, can be deduced from the double difference of binding energies (obtained directly from mass measurements) of the nucleus of interest and its neighbors. For odd-odd nuclei with $N$ neutrons and $Z$ protons, whose binding energy is $B(N,Z)$, the expression is \cite{VanIsacker1995}:

\begin{eqnarray}
\delta V_{pn}(N,Z)=&&[B(N,Z)-B(N-1,Z)]\nonumber\\
&&-[B(N,Z-1)-B(N-1,Z-1)].
\label{eq_dvpn_def}
\end{eqnarray}

Assuming that the nuclear core remains essentially unchanged in the four nuclei of Eq. \ref{eq_dvpn_def}, such a double difference largely cancels out the p-p and n-n pairing interactions and the mean field component of the binding energy, isolating the empirical p-n interaction strength \cite{Zhang1989,Cakirli2005,Cakirli2006,Stoitsov2007}. Moreover, $\delta$V$_{pn}$ can be used as a sensitive filter to study anomalies on the mass surface \cite{Cakirli2010b}.

It has been predicted that $\delta$V$_{pn}$ depends on the spatial overlap of wave functions of the last neutron and proton \cite{Brenner2006}. This prediction seems to be corroborated by experimental data around doubly-magic nuclei \cite{Chen2009,Zhang2018}, and especially by the dramatic increase of $\delta$V$_{pn}$ at N=Z nuclei, which is attributed to the manifestation of Wigner's SU(4) isospin and spin symmetry \cite{VanIsacker1995}.

Due to the relation to the overlap of the wave functions, it is expected that for N=Z nuclei (and in general), $\delta$V$_{pn}$ will decrease with increasing Z. This is because of three cumulative reasons. First, as Z increases, the nuclear radii increase, the average distance between the last proton and neutron increases, and their mutual interaction strength decreases even for nucleons occupying the same orbitals. Second, as Z increases, the Coulomb force causes the proton and neutron single-particle energies to differ even when they are in the same orbital \cite{Brenner2006}. Third, the spin-orbit term in the nuclear mean-field potential increases with mass, gradually breaking the spin part of Wigner's SU(4) symmetry \cite{VanIsacker1995}. 
The decrease with Z has been observed experimentally for all measured even-even nuclei \cite{Brenner2006} and odd-odd nuclei up to Z=29 \cite{Schury2007}.

\begin{figure*}
\centering
\includegraphics[width=1.0\textwidth]{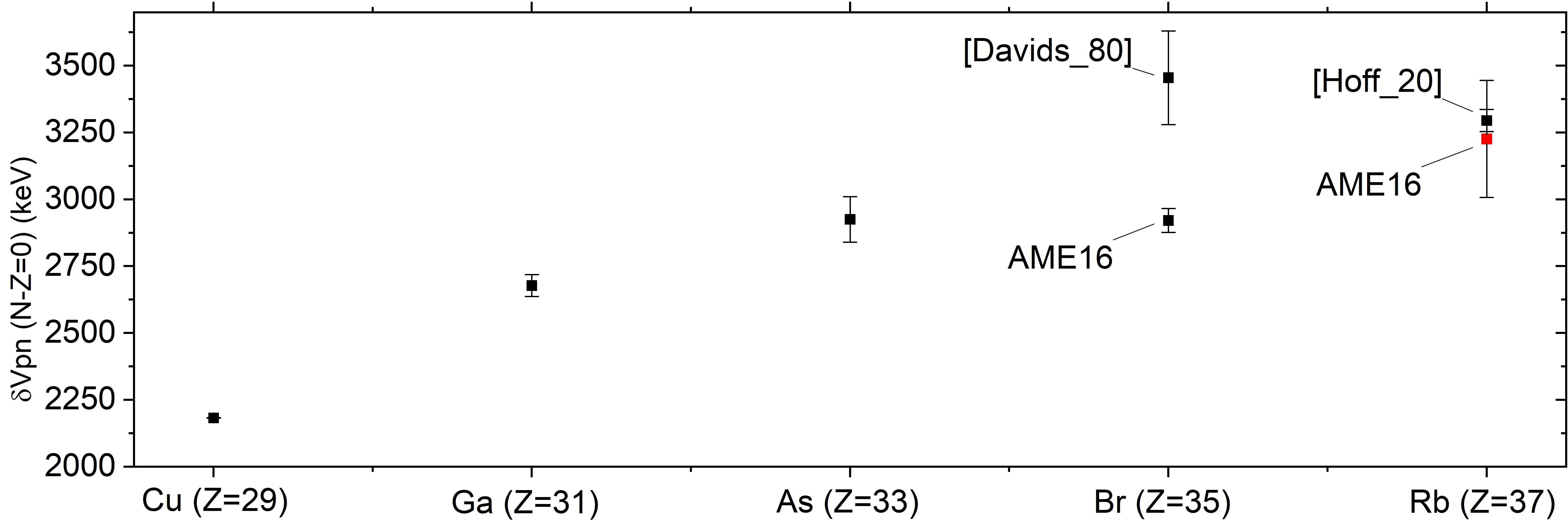} 
\caption{\label{fig:figure_dVpn0} Proton-neutron interaction strength as a function of Z for odd-odd N=Z nuclei from the nuclide $^{58}$Cu to the nuclide $^{74}$Rb. Black points are experimental data or AME16 \cite{Wang2017} evaluations based on experimental data. The red point for the nuclide $^{74}$Rb is the AME16 evaluation based on systematics. Hoff$\_$20 refers to \cite{Hoff2020} and Davids$\_$80 refers to \cite{Davids1980}.}
\end{figure*}

The authors of Ref. \cite{Schury2007} note that from Z=29 onwards, 'there is an indication of a trend toward the restrengthening of $\delta$V$_{pn}$'. We submit that based on a newer direct mass measurement of $^{65}$As \cite{Tu2011} and a very recent indirect mass measurement of $^{73}$Rb \cite{Hoff2020}, this restrengthening trend may be confirmed, as shown in  \ref{fig:figure_dVpn0}, which depicts the experimental data for N=Z odd-odd nuclei from Z=29 to Z=37.

For the nuclide $^{74}$Rb, we included in Fig. \ref{fig:figure_dVpn0} both the $\delta$V$_{pn}$ that is extracted from the AME16 \cite{Wang2017} mass value of $^{73}$Rb and that from Ref. \cite{Hoff2020}, to show how the 2020 measurement emphasizes the restrengthening trend. 
For the nuclide $^{70}$Br, we included in Fig. \ref{fig:figure_dVpn0} both the AME16 value that is based on the Penning trap measurement of \cite{Savory2009}, and the older indirect measurement based on positron endpoint energy \cite{Davids1980}, due to the approximate 500 keV discrepancy described above. 

The different values for the nuclide $^{70}$Br give systematically different local trends of $\delta$V$_{pn}$, either peaking at this nuclide \cite{Davids1980} or flattening out \cite{Savory2009} before continuing to increase towards the nuclide $^{74}$Rb. 
A new mass measurement of the nuclide $^{70}$Br and heavier nuclides in the N$\leq$Z region should resolve the question which is the correct trend.

For the mass measurement of such nuclei, which are mostly very short-lived (10's of ms) and with low production cross sections, an optimal device is the multiple-reflection time-of-flight mass spectrometer (MR-TOF-MS) \cite{Wollnik1990}, which is sensitive even to nuclei that are produced at rates as low as a few events per hour or day \cite{Plass2013b,Hornung2020}. 
The MR-TOF-MS has a unique combination of performance parameters: fast (cycle times of a few milliseconds), accurate (relative mass-measurement uncertainty down to the 10$^{-8}$ level), sensitive (only a few detected ions per nucleus are required for accurate mass determination), non-scanning (simultaneous measurement of many different nuclei) and can be used to spatially resolve isobars and even isomers \cite{Dickel2015,Ayet2019,Hornung2020}.

In this work, we present accurate MR-TOF-MS mass measurements of the nuclides $^{69}$As, $^{70,71}$Se and $^{71}$Br, all near the N=Z line. These results help to confirm trends of $\delta$V$_{pn}$ as a function of Z for odd-odd N-Z=2 and N-Z=4 nuclides in the region Z=29-37, which may provide hints on the expected N=Z $\delta$V$_{pn}$ behavior around the nuclide $^{70}$Br.

\section{Experimental Setup}
\label{sec:experimental}

The nuclei presented in this work were produced and spatially separated at relativistic energies via projectile fragmentation at the Fragment Separator (FRS) at GSI \cite{Geissel1992b}, and delivered to the FRS Ion Catcher \cite{Plass2013} where the ions were slowed down and thermalized in a gas-filled Cryogenic Stopping Cell (CSC) \cite{Ranjan2011,Purushothaman2013,Ranjan2015}. Subsequently, the ions were extracted and transported via a Radio Frequency Quadrupole (RFQ) beam-line \cite{Greiner2019, Haettner2018} to an MR-TOF-MS \cite{Plass2008,Dickel2015b}, to perform mass measurements and ion counting \cite{Ayet2019} with mass resolving powers at full width at half maximum (FWHM) of almost 1,000,000. We previously reached a resolving power above 1,000,000 for stable $^{133}$Cs ions with our MR-TOF-MS \cite{will2019,Beck2020}, and this is the first time we approach this value with unstable ions.

The production mechanism was fragmentation of a $^{124}$Xe primary beam impinging on a Be target of 4009 mg/cm$^2$ thickness at 982 MeV/u, with intensities up to 3$\times10^9$ ions per spill and a typical spill length of 2.5 seconds. The FRS was set up in achromatic mode from the target to the mid focal plane by using a wedge shaped aluminum degrader (141 mrad, 4005 mg/cm$^2$) at the focal plane F1. At the mid focal plane a secondary reaction target was installed. From the mid to the final focus the FRS was set to a dispersive mode. A new degrader system, including a variable angle degrader, was used at the final focal plane to range-bunch the ions. The measured range distribution after range bunching was about 50 mg/cm$^2$ \cite{Groef2020}. All nuclides in this work were measured with the same FRS setting, and only the degrader thickness in the final focal plane was changed. In addition to their accurate mass measurement in the MR-TOF-MS, the ions were also identified by the FRS particle ID system.

This enabled the parallel operation of two independent experiments with the same secondary beam - proton decaying nuclei were investigated with the EXPERT setup (see, e.g., \cite{Mukha2018}) installed after the secondary reaction target at the mid focal plane, and the longer-living nuclei that are described in this work were studied at the FRS Ion Catcher at the final focus of the FRS.

The CSC temperature during the experiment was 130 K, higher than the standard working temperature ($\sim$85 K), since the CSC had to be warmed up for repair just prior to the experiment, and the available cooling time was not sufficient for reaching the standard temperature.
The potential impact of a higher CSC temperature is an increase of the amount and variety of contaminants in the buffer gas, most of which are frozen at the standard working temperature.
Nevertheless, the total amount of contaminants in our acquired mass spectra was small, and their overall effect on our results was minute and under control.
The CSC was operated with an areal density of 3.3 mg/cm$^2$ and a mean extraction time of about 80 ms. 
\section{Data Analysis}
\label{sec:data_analysis}

We analyzed the data according to the procedure presented in \cite{Ayet2019}. Here we emphasize the parts that were particularly important for this work. The drifts of the time-of-flight data during the experiment were corrected by performing a time-resolved calibration using a well-known mass. The peak shape was obtained from a high-count reference and used for fitting the ion-of-interest. The analytical function describing the peaks is the Hyper-EMG \cite{Purushothaman2017}, and a weighted maximum likelihood estimate (wMLE) was used to fit this function to the un-binned data. All measured nuclei in this work were measured as singly-charged positive ions.

The relationship between time-of-flight and mass-to-charge is defined in equation \ref{eq:TOF_to_m},

\begin{equation} \label{eq:TOF_to_m}
    \frac{m}{q} = \frac{c(t_{\mathrm{exp}}-t_0)^2}{(1+N_{\mathrm{it}}b)^2}
\end{equation}

where $b = l_{\mathrm{it}}/l_{\mathrm{tfs}}$ ($l_{\mathrm{it}}$ is the path length for one turn in the analyzer and $l_{\mathrm{tfs}}$ is the path length from the injection trap to the detector), $c = 2U_{\mathrm{eff}}/l_{\mathrm{tfs}}^2$ ($U_{\mathrm{eff}}$ is the effective voltage), $t_{\mathrm{exp}}$ is the total measured time-of-flight that takes into account a time delay between the start signal and the real start of the ions ($t_0$), and $N_{\mathrm{it}}$ is the number of turns in the analyzer (see \cite{Ayet2019} for a more detailed description). 
The parameter $t_0$ was common for all the data analyzed in the context of this work and was determined to be 274 $\pm$ 2 ns.  

Mass values were obtained using calibrant ions that made the same and different turn numbers with respect to the ion-of-interest. We used four stable molecules that were present in the spectrum and underwent three different turn numbers in the analyzer, and whose relative mass accuracy $\delta m/m$ is in the $10^{-11}$ level (much better than the accuracy that we expect for the unstable nuclei of interest): $^{12}$C$_{5} \! ^{1}$H$_{10}$ (A=70) performing 847 turns, $^{12}$C$_{5} \! ^{1}$H$_{9}$ (A=69) performing 853 turns, and $^{14}$N$^{15}$N$^{40}$Ar (A=69) and $^{12}$C$^{19}$F$_3$ (A=69) each performing 854 turns. 
We always used $^{12}$C$^{19}$F$_3$ (the most abundant peak in the spectrum) together with two other molecules to perform the calibration for the mass measurement of the fourth molecule. 

In practice, we used $^{12}$C$^{19}$F$_3$ (A=69) together with a) $^{12}$C$_{5} \! ^{1}$H$_{10}$ (A=70) and $^{12}$C$_{5} \! ^{1}$H$_{9}$ to measure $^{14}$N$^{15}$N$^{40}$Ar (A=69) for the same-turn accuracy test, with b) $^{12}$C$_{5} \! ^{1}$H$_{10}$ and $^{14}$N$^{15}$N$^{40}$Ar (A=69) to measure $^{12}$C$_{5} \! ^{1}$H$_{9}$, and with c) $^{12}$C$_{5} \! ^{1}$H$_{9}$ and $^{14}$N$^{15}$N$^{40}$Ar (A=69) to measure $^{12}$C$_{5} \! ^{1}$H$_{10}$. 

The measurement with the same-turn calibration of $^{14}$N$^{15}$N$^{40}$Ar (A=69) showed a deviation from literature of (0.6 $\pm$ 1.1) keV, which represents a relative mass uncertainty 1.7$\times 10^{-8}$. 
For the multi-turn calibration, the measurement of $^{12}$C$_{5} \! ^{1}$H$_{10}$ and $^{12}$C$_{5} \! ^{1}$H$_{9}$ resulted in a deviation from literature of (-1.2 $\pm$ 2.9) keV ($\delta m/m$ of 4.4$\times 10^{-8}$) and (0.9 $\pm$ 2.9) keV ($\delta m/m$ of 4.5 $\times 10^{-8}$), respectively, with only about 300 events for each molecule.

The above analyses of the molecular masses show that both of our calibration procedures (same-turn and multi-turn) provide accuracy in the low $10^{-8}$ level.
We have reached in the past relative uncertainties of 6.0$\times 10^{-8}$ with our system, but then it was achieved with approximately 6000 events \cite{Ayet2019}. 
The relative mass uncertainty of 1.7 $\times 10^{-8}$ achieved for $^{14}$N$^{15}$N$^{40}$Ar (A=69) in this work is the most accurate measurement yet performed with an MR-TOF-MS, due to our unprecedented mass resolving power of almost 1,000,000 FWHM.
The uncertainty evaluation of the molecules and unstable nuclides of this work include all the systematic contributions that were presented in Ref. \cite{Ayet2019}.

\section{Results}
\label{sec:results}

\subsection{Mass Measurement of the nuclide $^{69}$As}
\label{sec:69As}

The mass of the nuclide $^{69}$As has been previously measured only indirectly, via the endpoint energy of its $\beta^+$ decay \cite{Boswell1970,Macdonald1977} and the endpoint energy of its mother nucleus (the nuclide $^{69}$Se) $\beta^+$ decay \cite{Macdonald1977}. 
The values from these three measurements (with uncertainties of 50 keV) present some inconsistency with each other, but nevertheless the value at AME16 is a weighted average of them, with a minor weight from the $\beta^+$ decay of the nuclide $^{69}$Se \cite{AME16}, and an overall uncertainty of 30 keV \cite{Wang2017}. 

\begin{figure*}
\centering
\resizebox{\textwidth}{!}{
\includegraphics{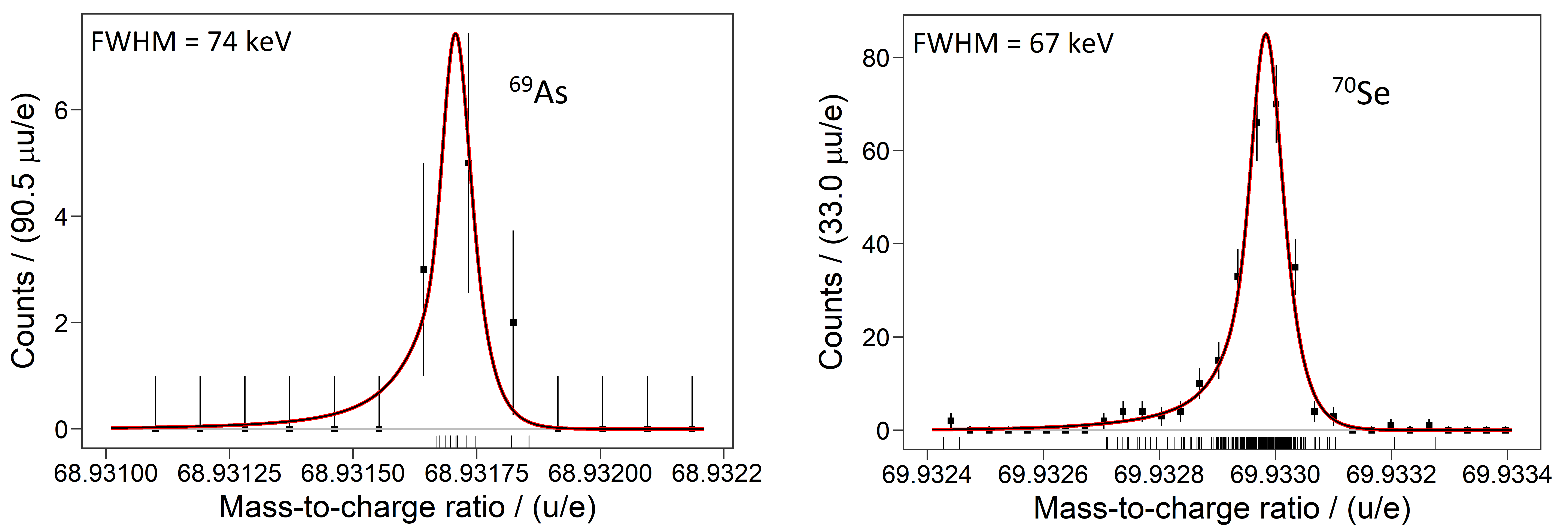}
}
\caption{Left panel: Measured mass-to-charge ratio spectrum of singly charged $^{69}$As nuclides. The square points represent the histogram of the un-binned data, which is shown as well in the lower part of the plot. A Hyper-EMG function with two exponentials on the left side and one exponential on the right side (Hyper-EMG(2,1)) with a FWHM of 74 keV (mass resolving power of 870,000 FWHM) was used for fitting the un-binned data ('rug' graph below the histogram), the shape parameters of which were obtained from the calibrant data (red line). Right panel: Same as on the left, for the nuclide $^{70}$Se, showing the results after the wMLE fit of a Hyper-EMG(2,1) with a FWHM of 67 keV (mass resolving power of 970,000 FWHM) to a set of data containing 256 counts of the nuclide $^{70}$Se.}
\label{fig:figure_masspeaks}       
\end{figure*}

In this work, we performed the first direct mass measurement of the nuclide $^{69}$As. The acquired data with the MR-TOF-MS contained only 10 events, highlighting the sensitivity of our system, as can be seen in the left side of Fig. \ref{fig:figure_masspeaks}. The total time-of-flight for this measurement was about 23.1 ms and the $^{69}$As ions performed 901 turns inside the analyzer.

Calibration was obtained with the molecular isobar $^{12}$C$^{19}$F$_3$ (A=69), which performed the same number of turns as the $^{69}$As ion. Some of these molecules are produced by ionization in the CSC and others from the electron impact source of the MR-TOF-MS (fragment of the molecule C$_{3}$F$_{8}$). 

The 10 events of the nuclide $^{69}$As were acquired in 40 minutes. When setting the degrader thickness in the FRS final focal plane to stop A=70 nuclides in CSC, we measured also the A=69 region for about 5 hours and did not observe any events in the mass region of the nuclide $^{69}$As. This provided confidence that the acquired 10 events are all $^{69}$As nuclides. Nevertheless, presence of unknown contamination in the peak could not be completely ruled out, and was thus taken into account in the uncertainty evaluation, according to the procedure described in Ref. \cite{Ayet2019}.

Due to the low number of events, the mass uncertainty of the nuclide $^{69}$As is dominated by statistics and a minor contribution from possible unknown contamination, because of the relatively low amount of unidentified peaks in the entire mass spectrum \cite{Ayet2019}. Its overall value is 22 keV, corresponding to a relative accuracy of $\delta m/m$ = 2.8$\times 10^{-7}$.

The mass-excess value and uncertainty of the nuclide $^{69}$As from this work are given in Table \ref{tab:results} and are compared to the AME16 value and previous indirect measurements in Fig. \ref{fig:figure_massres}. It can be seen in Fig. \ref{fig:figure_massres} that our result resolves the discrepancy between the previous indirect mass measurements of the nuclide $^{69}$As.

\subsection{Mass Measurement of the nuclide $^{70}$Se}
\label{sec:70Se}

The mass of the nuclide $^{70}$Se has been previously measured both indirectly and directly. The direct measurements have been performed with time-of-flight (GANIL \cite{Lima2002,Chartier1998}), Penning traps (ISOLTRAP at ISOLDE/CERN \cite{Herfurth2011} and LEBIT at NSCL \cite{Savory2009}) and isochronous mass measurements (ESR at GSI \cite{Hausmann2001}). 
The AME16 \cite{Wang2017} mass value and uncertainty (1.6 keV) are based mainly on the results from Ref. \cite{Savory2009}. 
The more recent Penning trap result is consistent with \cite{Savory2009} but with a larger uncertainty (17 keV) \cite{Herfurth2011}. The isochronous measurement result \cite{Hausmann2001} has a large uncertainty (70 keV) and deviates by 2$\sigma$ from the measurements with the Penning traps. The direct time-of-flight results are with even larger uncertainties, 130 keV \cite{Lima2002} and 460 keV \cite{Chartier1998}, and both are consistent with the Penning trap measurements.

In this work we performed the first mass measurement of the nuclide $^{70}$Se with an MR-TOF-MS. The total number of accumulated events is 485 with 2 different turn numbers (848 and 895) with a total TOF of about 21.9 and 23.1 ms, respectively. 

In the data sets where the $^{70}$Se ion underwent 848 turns, an isobaric molecule $^{13}$C$^{19}$F$_3$ (A=70) undergoing also 848 turns (produced partially in the CSC and partially from the electron impact source in the MR-TOF-MS) was used as a calibrant. 
In the measurement with 895 turns the ion of interest performed a different number of turns than the calibrant. Therefore, an accurate multi-turn calibration was performed by the determination of the parameter $c$ (see Eq. \ref{eq:TOF_to_m}) using other singly charged species performing different turn numbers, namely: $^{12}$C$_{5} \! ^{1}$H$_{10}$ (A=70) performing 847 turns, $^{12}$C$_{5} \! ^{1}$H$_{9}$ (A=69) performing 853 turns and $^{14}$N$^{15}$N$^{40}$Ar (A=69) and $^{12}$C$^{19}$F$_3$ (A=69) performing 854 turns. 

The parameter $b$ (see Eq.\ref{eq:TOF_to_m}) was determined in the different measurements from $^{12}$C$^{19}$F$_3$ (A=69) or $^{14}$N$^{15}$N$^{40}$Ar (A=69) performing 854 turns. Following the procedure described in \cite{Ayet2019}, an uncertainty of 2.6 keV was obtained, which corresponds to a relative accuracy $\delta m/m$ of 4.0$\times 10^{-8}$. A sample spectrum containing 256 counts is shown in Fig. \ref{fig:figure_masspeaks}.

The mass-excess value and uncertainty of the nuclide $^{70}$Se from this work are given in Table \ref{tab:results} and are compared together with previous measurements to the value given in AME16 \cite{Wang2017} in Fig. \ref{fig:figure_massres}. It can be seen in Fig. \ref{fig:figure_massres} that our result is consistent with previous high accuracy measurements and has a precision similar to the best one. 

Our relative uncertainty for the nuclide $^{70}$Se is similar to the best relative uncertainty recorded so far for an unstable nucleus by an MR-TOF-MS, $\delta m/m$ = 3.5$\times 10^{-8}$ for $^{65}$Ga in GARIS-II at RIKEN \cite{Kimura2018}. Notice that our level of uncertainty was obtained with almost a factor of 40 less events due to the much higher mass resolving power in our MR-TOF-MS. 

\subsection{Mass Measurement of the nuclide $^{71}$Se}
\label{sec:71Se}

The mass of the nuclide $^{71}$Se has been previously measured both indirectly and directly. The direct measurements have been performed with time-of-flight (GANIL \cite{Lima2002,Chartier1998}), Penning trap (ISOLTRAP at ISOLDE/CERN \cite{Herfurth2011}) and isochronous mass measurements (ESR at GSI \cite{Hausmann2001}). The AME16 \cite{Wang2017} mass value and uncertainty (2.8 keV) are based mainly on the results from Ref. \cite{Herfurth2011}. The isochronous measurement result \cite{Hausmann2001} has a large uncertainty (70 keV) and deviates by 1.3$\sigma$ from the Penning trap measurements. The direct time-of-flight results have even larger uncertainties, 112 keV \cite{Lima2002} and 317 keV \cite{Chartier1998}, and both are consistent with the Penning trap measurements (see Fig. \ref{fig:figure_massres} for a comparison with AME16).

The mass of the nuclide $^{71}$Se was measured with only 7 counts. The ions underwent 895 turns in about 23.1 ms. There was no same-turn calibration possible and the measurement was performed calculating the parameter $c$ (see Eq. \ref{eq:TOF_to_m}) from the multi-turn calibration used with $^{70}$Se and using $^{13}$C$^{19}$F$_3$ (A=70) as a precision calibrant, which performed 6 turns more than $^{71}$Se (901 turns). The uncertainty achieved was 23 keV or a relative mass accuracy $\delta m/m$ of 3.4$\times 10^{-7}$, where the main contributions are statistical error and a non-perfect correction of the TOF drifts. 

\subsection{Mass Measurement of the nuclide $^{71}$Br}
\label{sec:71Br}

\begin{figure*}
\centering
\resizebox{\textwidth}{!}{
\includegraphics{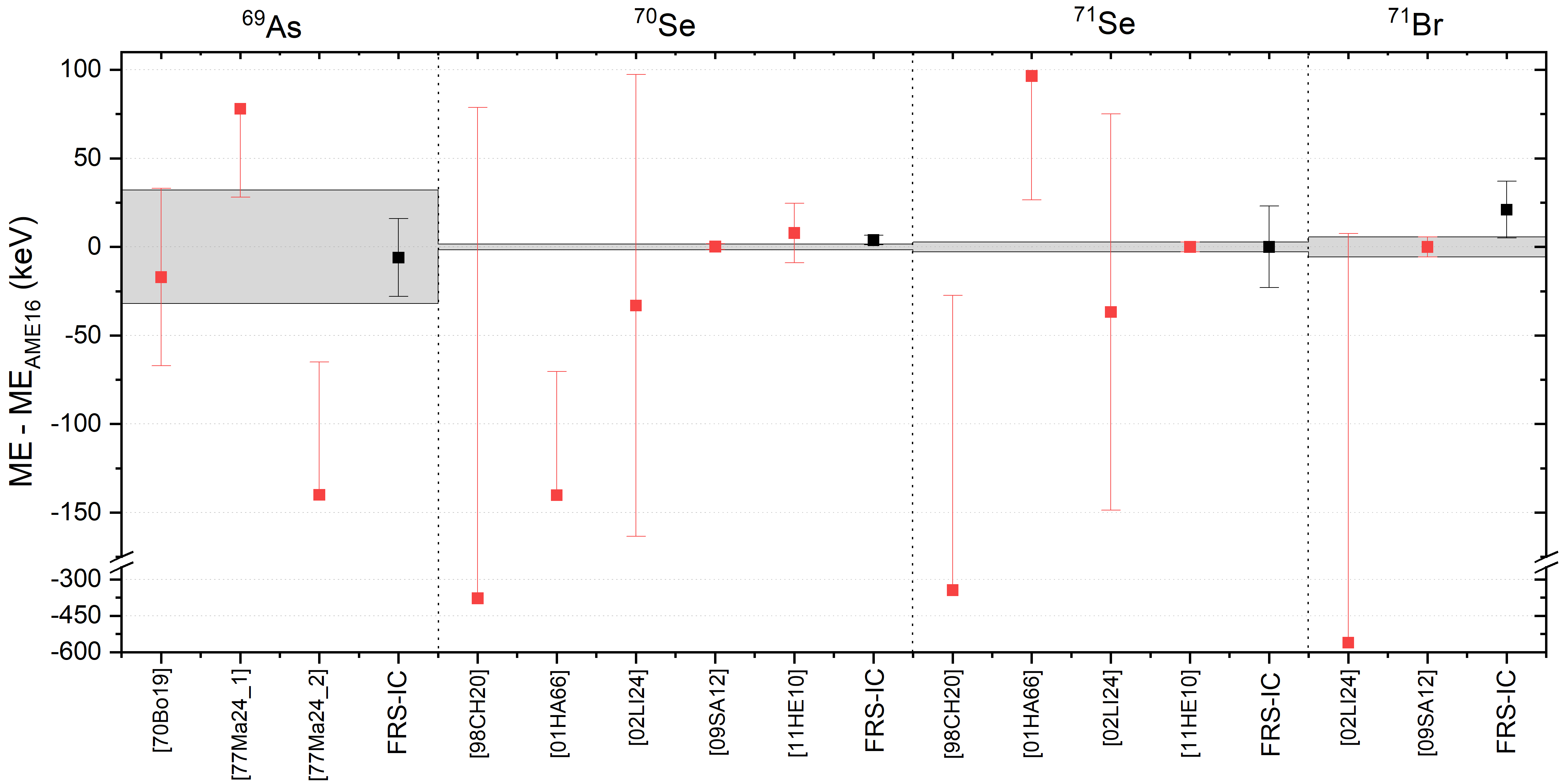}
}
\caption{Deviation of the measured mass-excess of this work (see Table \ref{tab:results}), denoted by 'FRS-IC', and previous ones from the values given in AME16 \cite{Wang2017}. [70Bo19] refers to \cite{Boswell1970}, [77Ma22\_1] and [77Ma22\_2] refer to \cite{Macdonald1977}, which includes the published endpoint energies of the decays $^{69}$As($\beta$+)$^{69}$Ge and $^{69}$Se($\beta$+)$^{69}$As, respectively, [98CH20] refers to \cite{Chartier1998}, [01HA66] refers to \cite{Hausmann2001}, [02LI24] refers to \cite{Lima2002}, [09SA12] refers to \cite{Savory2009} and [11HE10] refers to \cite{Herfurth2011}. The gray band around the horizontal axis represents the AME16 uncertainty. Breaks in the Y axis with different scale widths are included in order to be able to see high-accuracy data as well as low-accuracy data. Note that if one of the borders of the error bar is out of the displayed region, then it is not shown.}
\label{fig:figure_massres}       
\end{figure*}

Due to its closest location to the N=Z line, the mass of the nuclide $^{71}$Br has been previously measured only twice, by time-of-flight (GANIL \cite{Lima2002}) and a Penning trap (LEBIT at NSCL \cite{Savory2009}). The AME16 mass value and uncertainty (5.4 keV) are based solely on the results of Ref. \cite{Savory2009}. The time-of-flight result has a much larger uncertainty, 570 keV \cite{Lima2002}, which is consistent with the value from Ref. \cite{Savory2009}. The comparison can be seen in Fig. \ref{fig:figure_massres}. 

The measurement of the nuclide $^{71}$Br included a total of 19 counts. The ions underwent 895 turns in about 23.1 ms. This nuclide was measured together with the nuclide $^{71}$Se and the same calibration procedure was performed. The mass is in good agreement with the one presented in AME16 and presents an uncertainty of 16 keV or a relative mass accuracy $\delta m/m$ of 2.5$\times 10^{-7}$.

\begin{table*}
\caption{\label{tab:results}Results table of the different nuclides measured in this work and their comparison with AME16 \cite{Wang2017}. Half-lives were extracted from \cite{Audi2016}. The symbol $\dagger$ denotes a literature value based on indirect mass measurements. See section \ref{sec:results} for information about the mass references used in the measurement of each nuclide.}

\begin{tabular}{p{2cm} p{3.5cm} S[table-format=-5.1(5)] S[table-format=-5.1(1)] S[table-format=2.1(3)] c}

\toprule
\multicolumn{1}{c}{Nuclei} &  \multicolumn{1}{c}{Half-Life}  &  \multicolumn{1}{c}{ME$_\mathrm{FRS-IC}$ [keV] } & \multicolumn{1}{c}{\hspace{0.5cm}ME$_\mathrm{AME16}$  [keV]} & \multicolumn{1}{c}{\hspace{0.5cm} ME$_\mathrm{FRS-IC}$ - ME$_\mathrm{AME16}$ [keV]  } & \multicolumn{1}{c}{\hspace{0.25cm}Events\hspace{0.25cm}} \\

\hline  
\centering$^{69}$As & \centering15.2 $\pm$ 0.2 \si{\minute} & -63116 \pm 22   &  \multicolumn{1}{S[table-format=-5.1(1)]}{-63110 \pm 30 \si{\dagger}} & -6 \pm 37 & \multicolumn{1}{c}{10} \\
\centering$^{70}$Se & \centering41.1 $\pm$ 0.3 \si{\minute} & -61926.0 \pm 2.6  & -61929.9 \pm 1.6 & 3.9 \pm 3.0  & \multicolumn{1}{c}{485} \\
\centering$^{71}$Se & \centering4.74 $\pm$ 0.05 \si{\minute} & -63147 \pm 23   & -63146.5 \pm 2.8 & 0 \pm 23  & \multicolumn{1}{c}{7} \\
\centering$^{71}$Br & \centering21.4 $\pm$ 0.6 \si{\second} &  -56481 \pm 16 & -56502 \pm 5 & 21 \pm 17 & \multicolumn{1}{c}{19} \\
\hline \hline
\end{tabular}
\end{table*}    


\section{Discussion}
\label{sec:disc}

The proton-neutron interaction strength for N=Z nuclei, $\delta$V$_{pn}$(N=Z), is expected to decrease with increasing mass (see Section \ref{sec:intro} \cite{VanIsacker1995,Brenner2006}) whereas recent mass measurements show a restrengthening of this parameter in the range Z=29-37, as depicted in Fig. \ref{fig:figure_dVpn0}.
This could be explained by various physical phenomena, including the amount of overlap between wave functions of valence protons and neutrons \cite{Brenner2006,Chen2009,Zhang2018}, the growth of deformation in nuclei \cite{Cakirli2006,Bonatsos2013}, and the manifestation of Wigner's SU(4) symmetry at N=Z nuclei \cite{VanIsacker1995}.

\begin{figure*}
\centering
\resizebox{\textwidth}{!}{
\includegraphics{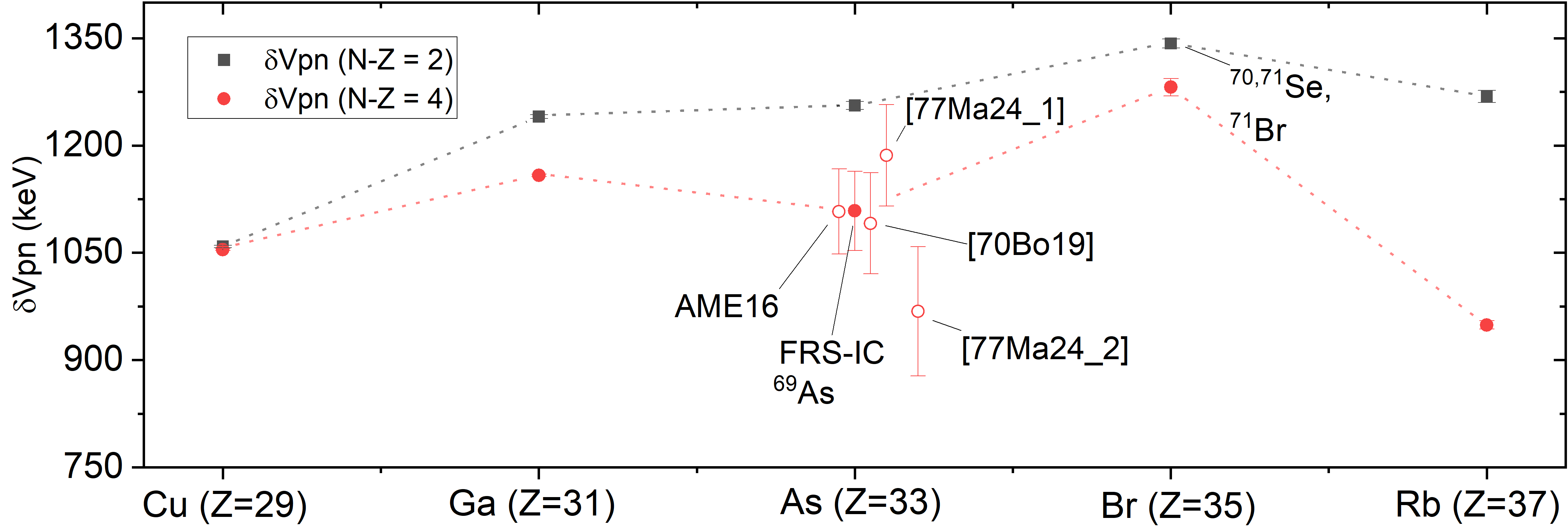}
}
\caption{Proton-neutron interaction strength as a function of Z for odd-odd N-Z=2 and N-Z=4 nuclei from Cu isotopes to Rb isotopes. Solid symbols are experimental data or AME16 evaluations \cite{Wang2017} based on experimental data. The open symbols for $\delta$V$_{pn}$(N-Z=4) of the As isotope are values based on the previous indirect mass measurements of the nuclide $^{69}$As. [70Bo19] refers to \cite{Boswell1970}, [77Ma22\_1] and [77Ma22\_2] refer to \cite{Macdonald1977} that includes the published endpoint energies of the decays $^{69}$As($\beta^+$)$^{69}$Ge and $^{69}$Se($\beta^+$)$^{69}$As, respectively. nuclide symbols near the points show the values which the measurement of this work had an impact on (the nuclide $^{69}$As) or confirmed (the nuclides $^{70,71}$Se, $^{71}$Br). Dotted lines are shown to guide the eye between $\delta$V$_{pn}$ points of the same N-Z value.}
\label{fig:figure_dVpn024}       
\end{figure*}

In order to discern which interpretation may be behind the surprising trend in $\delta$V$_{pn}$(N=Z), we applied the measured masses of this work and corresponding data from AME16 \cite{Wang2017} and primary literature \cite{Boswell1970,Macdonald1977} to calculate the quantity $\delta$V$_{pn}$ for N-Z=2 and N-Z=4 nuclides in the proton number range Z=29-37. The results are shown in Fig. \ref{fig:figure_dVpn024}.

From comparing Fig. \ref{fig:figure_dVpn024} to Fig. \ref{fig:figure_dVpn0}, it is observed that $\delta$V$_{pn}$(N=Z) is much higher (up to a factor of almost 3) than the values for N-Z=2 and N-Z=4, as expected \cite{VanIsacker1995}.
Our new direct measurement of the nuclide $^{69}$As confirms an essentially monotonic increase of $\delta$V$_{pn}$ staggering as a function of Z, as one moves away from N=Z towards N-Z=2 and N-Z=4. It is thus evident that the significant restrengthening of $\delta$V$_{pn}$ (approximately 1200 keV, see Fig. \ref{fig:figure_dVpn0}) is a sole characteristic of the N=Z nuclei. The N-Z=2 net increase is only about 200 keV in the same range, and at N-Z=4 there is staggering that ends with a net decrease of approximately 100 keV in the overall range.

This may imply that the restrengthening of $\delta$V$_{pn}$(N=Z) is related to partial restoration of a symmetry that is unique to these nuclei. 
As explained in Section \ref{sec:intro}, Wigner’s spin-isospin SU(4) symmetry is known to be broken in heavier nuclei \cite{VanIsacker1995,Brenner2006}. Nevertheless, it has been proposed that protons and neutrons occupying the pf-shell above the magic number 28 exhibit a pseudo-SU(4) symmetry \cite{VanIsacker1999}, which follows from the combined invariance in pseudospin and isospin. Pseudospin is defined for the pf-shell orbitals as a reduction by one unit of angular momentum, so they behave as pseudo sd-shell orbitals. Therefore, their pseudospin-orbit interaction is reduced accordingly, pseudo-SU(4) symmetry is restored, and one may conclude that this causes the increase of the $\delta$V$_{pn}$(N=Z) values.

The pseudo-SU(4) scheme has been used for a qualitative understanding of the observed Gamow-Teller $\beta$-decay of pf-shell nuclei \cite{VanIsacker1999}. The authors of Ref. \cite{VanIsacker1999} stated that pseudo-SU(4) could be tested via the trend of $\delta$V$_{pn}$(N=Z) values in the pf-shell. However, the needed mass measurements were lacking at that time. The experimental information that is presented here emphasizes the unique behavior of the N=Z nuclei in the pf-shell and can contribute to the test of the pseudo-SU(4) concept.

It is interesting to study also the detailed structure of $\delta$V$_{pn}$ as a function of Z for the N-Z=2 and N-Z=4 nuclei. The accurate measurements reveal the staggering and overall increase in $\delta$V$_{pn}$ from the Cu (Z=29) isotopes to the Br (Z=35) isotopes, and show the beginning of a decrease at the Rb (Z=37) isotopes. Since the $\delta$V$_{pn}$ values beyond the closed shell 28 seem to be influenced by the shell correction and by deformation changes in the four neighboring nuclei, these results may provide input to detailed calculations of nuclear deformation \cite{Cakirli2010b}.

Some detailed aspects of the $\delta$V$_{pn}$ trends of N-Z=4, N-Z=2 and N=Z nuclei may be similar. The most striking feature in the N-Z=4 and N-Z=2 trends may be the maximal value of $\delta$V$_{pn}$ at the Br isotopes. If one assumes that this structure is maintained also at N=Z, then Fig. \ref{fig:figure_dVpn0} indicates that the 'correct' value for the mass of the nuclide $^{70}Br$ is closer to the indirect mass measurement \cite{Davids1980} rather than the more recent Penning trap measurement \cite{Savory2009}. Such a suggested 'peak' at the nuclide $^{70}$Br may be connected to the observed transition from spherical to deformed nuclear shapes along the N=Z line at Z=35 \cite{Hasegawa2007}. Recall that the indirect mass result of $^{70}$Br \cite{Davids1980} also leads to a ${Ft}$-value that is consistent with the world average \cite{Hardy2015}.

\section{Summary and Conclusions}
\label{sec:conc}

In this work we present the mass measurement of four  very-neutron-deficient nuclides with the MR-TOF-MS at the FRS Ion Catcher at GSI, with an unprecedented mass resolving power of almost 1,000,000. This is a very important milestone for mass measurements of rare short-lived nuclides, since high resolving power is the only way to achieve the required uncertainties and resolve overlapping peaks when the number of accumulated events is inherently limited.
We performed the first direct mass measurement of the nuclide $^{69}$As, and the first MR-TOF-MS mass measurements of the nuclides $^{70,71}$Se and $^{71}$Br.

We carried out these measurements simultaneously with another experiment that used the same secondary beam to study proton decaying nuclei at the mid-focal plane of the FRS. This is an important demonstration of efficient use of rare ion beams that are currently in great demand at large accelerator facilities.

For the nuclide $^{70}$Se, we reached a relative mass uncertainty of $\delta m/m$ = 4.0$\times 10^{-8}$ with less than 500 events, similar to the best ever measurement of an unstable nucleus with an MR-TOF-MS \cite{Kimura2018}. 
For the other three nuclides, we reached a relative mass uncertainty of about $2\times 10^{-7}$ with less than 20 events.
All our results are in good agreement with the AME16 values.

We measured the mass of the molecule $^{14}$N$^{15}$N$^{40}$Ar (A=69) with a relative accuracy of $\delta m/m$ = 1.7$\times 10^{-8}$, the highest yet achieved with an MR-TOF-MS.

Our first direct measurement of the nuclide $^{69}$As has a better uncertainty than both previous indirect measurements and the AME16 evaluation. 
Therefore, our results significantly increase the confidence in all four AME16 values, and the consequential conclusions of the nuclear structure and astrophysics models that rely on them.

Our results impact (the nuclide $^{69}$As) and confirm (the nuclides $^{70,71}$Se, $^{71}$Br) interesting trends in the proton-neutron interaction strength $\delta$V$_{pn}$ for N-Z=2 and N-Z=4 in the Z=29-37 range, which singled out the N=Z nuclei in this region as exhibiting the restrengthening of $\delta$V$_{pn}$ with increasing Z, in contrast with the expected behavior. Further, the maximal $\delta$V$_{pn}$ value at the Br isotopes with N-Z=2 and N-Z=4 may indicate that the indirect mass measurement \cite{Davids1980} of the nuclide $^{70}$Br may be more consistent with the regional $\delta$V$_{pn}$ trend than the Penning trap measurement \cite{Savory2009}. This gives another important incentive to re-measure the mass of this nuclide, in addition to the apparent deviation of its Q$_{EC}$ from the value that is consistent with the ${Ft}$ world average \cite{Hardy2015,Hardy2020}. From the technical point of view, the high accuracy achieved in the measurements of this work with a small number of events ensure that the MR-TOF-MS at the FRS Ion Catcher at GSI is ready for accurate measurements of the masses of rare N$\leq$Z nuclides in the A$\approx$70 region. In particular, our mass measurement of the nuclide $^{71}$Br shows that our system is able to extract the specially reactive element bromine, and thus our path to measure the nuclide $^{70}$Br and heavier N=Z nuclides is open and will be realized.

\section*{Acknowledgments}

The results presented here were obtained in experiment S459+, which was performed at the FRS at the GSI Helmholtzzentrum f{\"u}r Schwerionenforschung, Darmstadt (Germany) in the framework of FAIR Phase-0. This work was carried out within the scientific program of the Super-FRS Experiment Collaboration.
This work was supported by the German Federal Ministry for Education and Research (BMBF) under contracts no.\ 05P12RGFN8, 05P16RGFN1 and 05P19RGFN1, Justus-Liebig-Universit{\"a}t Gie{\ss}en and GSI under the JLU-GSI strategic Helmholtz partnership agreement, HGS-HIRe, the Hessian Ministry for Science and Art (HMWK) through the LOEWE Center HICforFAIR, Polish National Science Centre (2016/21/B/ST2/01227), UKRI STFC grant No. ST/P004008/1, Ministry of Education, Youth and Sports, Czech Republic (Projects No. LTT17003 and LM2018112), and by the Israel Ministry of Energy, Research Grant No. 217-11-023.
This project has received funding from the European Union’s Horizon 2020 research and innovation programme under grant agreements No. 654002 and No. 771036 (ERC CoG MAIDEN). The ELI-NP team acknowledges the support from the Extreme Light Infrastructure Nuclear Physics (ELI-NP) Phase II, a project co-financed by the Romanian Government and the European Union through the European Regional Development Fund - the Competitiveness Operational Programme (1/07.07.2016, COP, ID 1334).
One of the authors (NK) would like to acknowledge the support from the ExtreMe Matter Institute EMMI at the GSI Helmholtzzentrum f{\"u}r Schwerionenforschung, Darmstadt (Germany).

\bibliographystyle{apsrev4-1}
\bibliography{References.bib}

\begin{thebibliography}{54}%
\makeatletter
\providecommand \@ifxundefined [1]{%
 \@ifx{#1\undefined}
}%
\providecommand \@ifnum [1]{%
 \ifnum #1\expandafter \@firstoftwo
 \else \expandafter \@secondoftwo
 \fi
}%
\providecommand \@ifx [1]{%
 \ifx #1\expandafter \@firstoftwo
 \else \expandafter \@secondoftwo
 \fi
}%
\providecommand \natexlab [1]{#1}%
\providecommand \enquote  [1]{``#1''}%
\providecommand \bibnamefont  [1]{#1}%
\providecommand \bibfnamefont [1]{#1}%
\providecommand \citenamefont [1]{#1}%
\providecommand \href@noop [0]{\@secondoftwo}%
\providecommand \href [0]{\begingroup \@sanitize@url \@href}%
\providecommand \@href[1]{\@@startlink{#1}\@@href}%
\providecommand \@@href[1]{\endgroup#1\@@endlink}%
\providecommand \@sanitize@url [0]{\catcode `\\12\catcode `\$12\catcode
  `\&12\catcode `\#12\catcode `\^12\catcode `\_12\catcode `\%12\relax}%
\providecommand \@@startlink[1]{}%
\providecommand \@@endlink[0]{}%
\providecommand \url  [0]{\begingroup\@sanitize@url \@url }%
\providecommand \@url [1]{\endgroup\@href {#1}{\urlprefix }}%
\providecommand \urlprefix  [0]{URL }%
\providecommand \Eprint [0]{\href }%
\providecommand \doibase [0]{http://dx.doi.org/}%
\providecommand \selectlanguage [0]{\@gobble}%
\providecommand \bibinfo  [0]{\@secondoftwo}%
\providecommand \bibfield  [0]{\@secondoftwo}%
\providecommand \translation [1]{[#1]}%
\providecommand \BibitemOpen [0]{}%
\providecommand \bibitemStop [0]{}%
\providecommand \bibitemNoStop [0]{.\EOS\space}%
\providecommand \EOS [0]{\spacefactor3000\relax}%
\providecommand \BibitemShut  [1]{\csname bibitem#1\endcsname}%
\let\auto@bib@innerbib\@empty
\bibitem [{\citenamefont {Dilling}\ \emph {et~al.}(2018)\citenamefont
  {Dilling}, \citenamefont {Blaum}, \citenamefont {Brodeur},\ and\
  \citenamefont {Eliseev}}]{Dilling2018}%
  \BibitemOpen
  \bibfield  {author} {\bibinfo {author} {\bibfnamefont {J.}~\bibnamefont
  {Dilling}}, \bibinfo {author} {\bibfnamefont {K.}~\bibnamefont {Blaum}},
  \bibinfo {author} {\bibfnamefont {M.}~\bibnamefont {Brodeur}}, \ and\
  \bibinfo {author} {\bibfnamefont {S.}~\bibnamefont {Eliseev}},\ }\href
  {\doibase 10.1146/annurev-nucl-102711-094939} {\bibfield  {journal} {\bibinfo
   {journal} {Annu. Rev. Nucl. Part. Sci.}\ }\textbf {\bibinfo {volume} {68}},\
  \bibinfo {pages} {45} (\bibinfo {year} {2018})},\ \Eprint
  {http://arxiv.org/abs/https://doi.org/10.1146/annurev-nucl-102711-094939}
  {https://doi.org/10.1146/annurev-nucl-102711-094939} \BibitemShut {NoStop}%
\bibitem [{\citenamefont {Zhang}\ \emph {et~al.}(2016)\citenamefont {Zhang},
  \citenamefont {Litvinov}, \citenamefont {Uesaka},\ and\ \citenamefont
  {Xu}}]{zhang2016}%
  \BibitemOpen
  \bibfield  {author} {\bibinfo {author} {\bibfnamefont {Y.~H.}\ \bibnamefont
  {Zhang}}, \bibinfo {author} {\bibfnamefont {Y.~A.}\ \bibnamefont {Litvinov}},
  \bibinfo {author} {\bibfnamefont {T.}~\bibnamefont {Uesaka}}, \ and\ \bibinfo
  {author} {\bibfnamefont {H.~S.}\ \bibnamefont {Xu}},\ }\href@noop {}
  {\bibfield  {journal} {\bibinfo  {journal} {{Phys. Scr.}}\ }\textbf {\bibinfo
  {volume} {{91}}},\ \bibinfo {pages} {073022} (\bibinfo {year}
  {2016})}\BibitemShut {NoStop}%
\bibitem [{\citenamefont {Schatz}(2013)}]{Schatz2013}%
  \BibitemOpen
  \bibfield  {author} {\bibinfo {author} {\bibfnamefont {H.}~\bibnamefont
  {Schatz}},\ }\href@noop {} {\bibfield  {journal} {\bibinfo  {journal} {Int.
  J. Mass Spectrom.}\ }\textbf {\bibinfo {volume} {349 - 350}},\ \bibinfo
  {pages} {181} (\bibinfo {year} {2013})}\BibitemShut {NoStop}%
\bibitem [{\citenamefont {Schatz}\ and\ \citenamefont
  {Ong}(2017)}]{Schatz2017}%
  \BibitemOpen
  \bibfield  {author} {\bibinfo {author} {\bibfnamefont {H.}~\bibnamefont
  {Schatz}}\ and\ \bibinfo {author} {\bibfnamefont {W.-J.}\ \bibnamefont
  {Ong}},\ }\href {\doibase 10.3847/1538-4357/aa7de9} {\bibfield  {journal}
  {\bibinfo  {journal} {Astrophys. J.}\ }\textbf {\bibinfo {volume} {844}},\
  \bibinfo {pages} {139} (\bibinfo {year} {2017})}\BibitemShut {NoStop}%
\bibitem [{\citenamefont {Duy}\ \emph {et~al.}(2020)\citenamefont {Duy},
  \citenamefont {Ho},\ and\ \citenamefont {Uyen}}]{Duy2020}%
  \BibitemOpen
  \bibfield  {author} {\bibinfo {author} {\bibfnamefont {N.~N.}\ \bibnamefont
  {Duy}}, \bibinfo {author} {\bibfnamefont {P.-T.}\ \bibnamefont {Ho}}, \ and\
  \bibinfo {author} {\bibfnamefont {N.~K.}\ \bibnamefont {Uyen}},\ }\href
  {\doibase 10.3938/jkps.76.881} {\bibfield  {journal} {\bibinfo  {journal} {J.
  Korean Phys. Soc.}\ }\textbf {\bibinfo {volume} {76}},\ \bibinfo {pages}
  {881} (\bibinfo {year} {2020})}\BibitemShut {NoStop}%
\bibitem [{\citenamefont {Hardy}\ and\ \citenamefont
  {Towner}(2015)}]{Hardy2015}%
  \BibitemOpen
  \bibfield  {author} {\bibinfo {author} {\bibfnamefont {J.~C.}\ \bibnamefont
  {Hardy}}\ and\ \bibinfo {author} {\bibfnamefont {I.~S.}\ \bibnamefont
  {Towner}},\ }\href {\doibase 10.1103/PhysRevC.91.025501} {\bibfield
  {journal} {\bibinfo  {journal} {Phys. Rev. C}\ }\textbf {\bibinfo {volume}
  {91}},\ \bibinfo {pages} {025501} (\bibinfo {year} {2015})}\BibitemShut
  {NoStop}%
\bibitem [{\citenamefont {Hardy}\ and\ \citenamefont
  {Towner}(2020)}]{Hardy2020}%
  \BibitemOpen
  \bibfield  {author} {\bibinfo {author} {\bibfnamefont {J.~C.}\ \bibnamefont
  {Hardy}}\ and\ \bibinfo {author} {\bibfnamefont {I.~S.}\ \bibnamefont
  {Towner}},\ }\href {\doibase 10.1103/PhysRevC.102.045501} {\bibfield
  {journal} {\bibinfo  {journal} {Phys. Rev. C}\ }\textbf {\bibinfo {volume}
  {102}},\ \bibinfo {pages} {045501} (\bibinfo {year} {2020})}\BibitemShut
  {NoStop}%
\bibitem [{\citenamefont {Morales}\ \emph {et~al.}(2017)\citenamefont
  {Morales}, \citenamefont {Algora}, \citenamefont {Rubio}, \citenamefont
  {Kaneko}, \citenamefont {Nishimura}, \citenamefont {Aguilera}, \citenamefont
  {Orrigo}, \citenamefont {Molina}, \citenamefont {de~Angelis}, \citenamefont
  {Recchia}, \citenamefont {Kiss}, \citenamefont {Phong}, \citenamefont {Wu},
  \citenamefont {Nishimura}, \citenamefont {Oikawa}, \citenamefont {Goigoux},
  \citenamefont {Giovinazzo}, \citenamefont {Ascher}, \citenamefont {Agramunt},
  \citenamefont {Ahn}, \citenamefont {Baba}, \citenamefont {Blank},
  \citenamefont {Borcea}, \citenamefont {Boso}, \citenamefont {Davies},
  \citenamefont {Diel}, \citenamefont {Dombr\'adi}, \citenamefont {Doornenbal},
  \citenamefont {Eberth}, \citenamefont {de~France}, \citenamefont {Fujita},
  \citenamefont {Fukuda}, \citenamefont {Ganioglu}, \citenamefont {Gelletly},
  \citenamefont {Gerbaux}, \citenamefont {Gr\'evy}, \citenamefont {Guadilla},
  \citenamefont {Inabe}, \citenamefont {Isobe}, \citenamefont {Kojouharov},
  \citenamefont {Korten}, \citenamefont {Kubo}, \citenamefont {Kubono},
  \citenamefont {Kurtuki\'an~Nieto}, \citenamefont {Kurz}, \citenamefont {Lee},
  \citenamefont {Lenzi}, \citenamefont {Liu}, \citenamefont {Lokotko},
  \citenamefont {Lubos}, \citenamefont {Magron}, \citenamefont
  {Montaner-Piz\'a}, \citenamefont {Napoli}, \citenamefont {Sakurai},
  \citenamefont {Schaffner}, \citenamefont {Shimizu}, \citenamefont {Sidong},
  \citenamefont {S\"oderstr\"om}, \citenamefont {Sumikama}, \citenamefont
  {Suzuki}, \citenamefont {Takeda}, \citenamefont {Takei}, \citenamefont
  {Tanaka},\ and\ \citenamefont {Yagi}}]{Morales2017}%
  \BibitemOpen
  \bibfield  {author} {\bibinfo {author} {\bibfnamefont {A.~I.}\ \bibnamefont
  {Morales}}, \bibinfo {author} {\bibfnamefont {A.}~\bibnamefont {Algora}},
  \bibinfo {author} {\bibfnamefont {B.}~\bibnamefont {Rubio}}, \bibinfo
  {author} {\bibfnamefont {K.}~\bibnamefont {Kaneko}}, \bibinfo {author}
  {\bibfnamefont {S.}~\bibnamefont {Nishimura}}, \bibinfo {author}
  {\bibfnamefont {P.}~\bibnamefont {Aguilera}}, \bibinfo {author}
  {\bibfnamefont {S.~E.~A.}\ \bibnamefont {Orrigo}}, \bibinfo {author}
  {\bibfnamefont {F.}~\bibnamefont {Molina}}, \bibinfo {author} {\bibfnamefont
  {G.}~\bibnamefont {de~Angelis}}, \bibinfo {author} {\bibfnamefont
  {F.}~\bibnamefont {Recchia}}, \bibinfo {author} {\bibfnamefont
  {G.}~\bibnamefont {Kiss}}, \bibinfo {author} {\bibfnamefont {V.~H.}\
  \bibnamefont {Phong}}, \bibinfo {author} {\bibfnamefont {J.}~\bibnamefont
  {Wu}}, \bibinfo {author} {\bibfnamefont {D.}~\bibnamefont {Nishimura}},
  \bibinfo {author} {\bibfnamefont {H.}~\bibnamefont {Oikawa}}, \bibinfo
  {author} {\bibfnamefont {T.}~\bibnamefont {Goigoux}}, \bibinfo {author}
  {\bibfnamefont {J.}~\bibnamefont {Giovinazzo}}, \bibinfo {author}
  {\bibfnamefont {P.}~\bibnamefont {Ascher}}, \bibinfo {author} {\bibfnamefont
  {J.}~\bibnamefont {Agramunt}}, \bibinfo {author} {\bibfnamefont {D.~S.}\
  \bibnamefont {Ahn}}, \bibinfo {author} {\bibfnamefont {H.}~\bibnamefont
  {Baba}}, \bibinfo {author} {\bibfnamefont {B.}~\bibnamefont {Blank}},
  \bibinfo {author} {\bibfnamefont {C.}~\bibnamefont {Borcea}}, \bibinfo
  {author} {\bibfnamefont {A.}~\bibnamefont {Boso}}, \bibinfo {author}
  {\bibfnamefont {P.}~\bibnamefont {Davies}}, \bibinfo {author} {\bibfnamefont
  {F.}~\bibnamefont {Diel}}, \bibinfo {author} {\bibfnamefont {Z.}~\bibnamefont
  {Dombr\'adi}}, \bibinfo {author} {\bibfnamefont {P.}~\bibnamefont
  {Doornenbal}}, \bibinfo {author} {\bibfnamefont {J.}~\bibnamefont {Eberth}},
  \bibinfo {author} {\bibfnamefont {G.}~\bibnamefont {de~France}}, \bibinfo
  {author} {\bibfnamefont {Y.}~\bibnamefont {Fujita}}, \bibinfo {author}
  {\bibfnamefont {N.}~\bibnamefont {Fukuda}}, \bibinfo {author} {\bibfnamefont
  {E.}~\bibnamefont {Ganioglu}}, \bibinfo {author} {\bibfnamefont
  {W.}~\bibnamefont {Gelletly}}, \bibinfo {author} {\bibfnamefont
  {M.}~\bibnamefont {Gerbaux}}, \bibinfo {author} {\bibfnamefont
  {S.}~\bibnamefont {Gr\'evy}}, \bibinfo {author} {\bibfnamefont
  {V.}~\bibnamefont {Guadilla}}, \bibinfo {author} {\bibfnamefont
  {N.}~\bibnamefont {Inabe}}, \bibinfo {author} {\bibfnamefont
  {T.}~\bibnamefont {Isobe}}, \bibinfo {author} {\bibfnamefont
  {I.}~\bibnamefont {Kojouharov}}, \bibinfo {author} {\bibfnamefont
  {W.}~\bibnamefont {Korten}}, \bibinfo {author} {\bibfnamefont
  {T.}~\bibnamefont {Kubo}}, \bibinfo {author} {\bibfnamefont {S.}~\bibnamefont
  {Kubono}}, \bibinfo {author} {\bibfnamefont {T.}~\bibnamefont
  {Kurtuki\'an~Nieto}}, \bibinfo {author} {\bibfnamefont {N.}~\bibnamefont
  {Kurz}}, \bibinfo {author} {\bibfnamefont {J.}~\bibnamefont {Lee}}, \bibinfo
  {author} {\bibfnamefont {S.}~\bibnamefont {Lenzi}}, \bibinfo {author}
  {\bibfnamefont {J.}~\bibnamefont {Liu}}, \bibinfo {author} {\bibfnamefont
  {T.}~\bibnamefont {Lokotko}}, \bibinfo {author} {\bibfnamefont
  {D.}~\bibnamefont {Lubos}}, \bibinfo {author} {\bibfnamefont
  {C.}~\bibnamefont {Magron}}, \bibinfo {author} {\bibfnamefont
  {A.}~\bibnamefont {Montaner-Piz\'a}}, \bibinfo {author} {\bibfnamefont
  {D.~R.}\ \bibnamefont {Napoli}}, \bibinfo {author} {\bibfnamefont
  {H.}~\bibnamefont {Sakurai}}, \bibinfo {author} {\bibfnamefont
  {H.}~\bibnamefont {Schaffner}}, \bibinfo {author} {\bibfnamefont
  {Y.}~\bibnamefont {Shimizu}}, \bibinfo {author} {\bibfnamefont
  {C.}~\bibnamefont {Sidong}}, \bibinfo {author} {\bibfnamefont {P.-A.}\
  \bibnamefont {S\"oderstr\"om}}, \bibinfo {author} {\bibfnamefont
  {T.}~\bibnamefont {Sumikama}}, \bibinfo {author} {\bibfnamefont
  {H.}~\bibnamefont {Suzuki}}, \bibinfo {author} {\bibfnamefont
  {H.}~\bibnamefont {Takeda}}, \bibinfo {author} {\bibfnamefont
  {Y.}~\bibnamefont {Takei}}, \bibinfo {author} {\bibfnamefont
  {M.}~\bibnamefont {Tanaka}}, \ and\ \bibinfo {author} {\bibfnamefont
  {S.}~\bibnamefont {Yagi}},\ }\href {\doibase 10.1103/PhysRevC.95.064327}
  {\bibfield  {journal} {\bibinfo  {journal} {Phys. Rev. C}\ }\textbf {\bibinfo
  {volume} {95}},\ \bibinfo {pages} {064327} (\bibinfo {year}
  {2017})}\BibitemShut {NoStop}%
\bibitem [{\citenamefont {Savory}\ \emph {et~al.}(2009)\citenamefont {Savory},
  \citenamefont {Schury}, \citenamefont {Bachelet}, \citenamefont {Block},
  \citenamefont {Bollen}, \citenamefont {Facina}, \citenamefont {Folden},
  \citenamefont {Gu\'enaut}, \citenamefont {Kwan}, \citenamefont {Kwiatkowski},
  \citenamefont {Morrissey}, \citenamefont {Pang}, \citenamefont {Prinke},
  \citenamefont {Ringle}, \citenamefont {Schatz}, \citenamefont {Schwarz},\
  and\ \citenamefont {Sumithrarachchi}}]{Savory2009}%
  \BibitemOpen
  \bibfield  {author} {\bibinfo {author} {\bibfnamefont {J.}~\bibnamefont
  {Savory}}, \bibinfo {author} {\bibfnamefont {P.}~\bibnamefont {Schury}},
  \bibinfo {author} {\bibfnamefont {C.}~\bibnamefont {Bachelet}}, \bibinfo
  {author} {\bibfnamefont {M.}~\bibnamefont {Block}}, \bibinfo {author}
  {\bibfnamefont {G.}~\bibnamefont {Bollen}}, \bibinfo {author} {\bibfnamefont
  {M.}~\bibnamefont {Facina}}, \bibinfo {author} {\bibfnamefont {C.~M.}\
  \bibnamefont {Folden}}, \bibinfo {author} {\bibfnamefont {C.}~\bibnamefont
  {Gu\'enaut}}, \bibinfo {author} {\bibfnamefont {E.}~\bibnamefont {Kwan}},
  \bibinfo {author} {\bibfnamefont {A.~A.}\ \bibnamefont {Kwiatkowski}},
  \bibinfo {author} {\bibfnamefont {D.~J.}\ \bibnamefont {Morrissey}}, \bibinfo
  {author} {\bibfnamefont {G.~K.}\ \bibnamefont {Pang}}, \bibinfo {author}
  {\bibfnamefont {A.}~\bibnamefont {Prinke}}, \bibinfo {author} {\bibfnamefont
  {R.}~\bibnamefont {Ringle}}, \bibinfo {author} {\bibfnamefont
  {H.}~\bibnamefont {Schatz}}, \bibinfo {author} {\bibfnamefont
  {S.}~\bibnamefont {Schwarz}}, \ and\ \bibinfo {author} {\bibfnamefont
  {C.~S.}\ \bibnamefont {Sumithrarachchi}},\ }\href {\doibase
  10.1103/PhysRevLett.102.132501} {\bibfield  {journal} {\bibinfo  {journal}
  {Phys. Rev. Lett.}\ }\textbf {\bibinfo {volume} {102}},\ \bibinfo {pages}
  {132501} (\bibinfo {year} {2009})}\BibitemShut {NoStop}%
\bibitem [{\citenamefont {Davids}(1980)}]{Davids1980}%
  \BibitemOpen
  \bibfield  {author} {\bibinfo {author} {\bibfnamefont {C.~N.}\ \bibnamefont
  {Davids}},\ }\enquote {\bibinfo {title} {Masses of new isotopes in the
  $\mathit{pf}$-shell},}\ in\ \href@noop {} {\emph {\bibinfo {booktitle} {6th
  International Conference on Atomic Masses, East Lansing, Mich., 1979 - Atomic
  Masses and Fundamental Constants 6}}}\ (\bibinfo  {publisher} {Plenum},\
  \bibinfo {address} {New York},\ \bibinfo {year} {1980})\ p.\ \bibinfo {pages}
  {419},\ \bibinfo {note} {edited by J. A. Nolen and W. Benenson}\BibitemShut
  {NoStop}%
\bibitem [{\citenamefont {Van~Isacker}\ \emph {et~al.}(1995)\citenamefont
  {Van~Isacker}, \citenamefont {Warner},\ and\ \citenamefont
  {Brenner}}]{VanIsacker1995}%
  \BibitemOpen
  \bibfield  {author} {\bibinfo {author} {\bibfnamefont {P.}~\bibnamefont
  {Van~Isacker}}, \bibinfo {author} {\bibfnamefont {D.~D.}\ \bibnamefont
  {Warner}}, \ and\ \bibinfo {author} {\bibfnamefont {D.~S.}\ \bibnamefont
  {Brenner}},\ }\href {\doibase 10.1103/PhysRevLett.74.4607} {\bibfield
  {journal} {\bibinfo  {journal} {Phys. Rev. Lett.}\ }\textbf {\bibinfo
  {volume} {74}},\ \bibinfo {pages} {4607} (\bibinfo {year}
  {1995})}\BibitemShut {NoStop}%
\bibitem [{\citenamefont {Zhang}\ \emph {et~al.}(1989)\citenamefont {Zhang},
  \citenamefont {Casten},\ and\ \citenamefont {Brenner}}]{Zhang1989}%
  \BibitemOpen
  \bibfield  {author} {\bibinfo {author} {\bibfnamefont {J.-Y.}\ \bibnamefont
  {Zhang}}, \bibinfo {author} {\bibfnamefont {R.}~\bibnamefont {Casten}}, \
  and\ \bibinfo {author} {\bibfnamefont {D.}~\bibnamefont {Brenner}},\ }\href
  {\doibase https://doi.org/10.1016/0370-2693(89)91273-2} {\bibfield  {journal}
  {\bibinfo  {journal} {Physics Letters B}\ }\textbf {\bibinfo {volume}
  {227}},\ \bibinfo {pages} {1 } (\bibinfo {year} {1989})}\BibitemShut
  {NoStop}%
\bibitem [{\citenamefont {Cakirli}\ \emph {et~al.}(2005)\citenamefont
  {Cakirli}, \citenamefont {Brenner}, \citenamefont {Casten},\ and\
  \citenamefont {Millman}}]{Cakirli2005}%
  \BibitemOpen
  \bibfield  {author} {\bibinfo {author} {\bibfnamefont {R.~B.}\ \bibnamefont
  {Cakirli}}, \bibinfo {author} {\bibfnamefont {D.~S.}\ \bibnamefont
  {Brenner}}, \bibinfo {author} {\bibfnamefont {R.~F.}\ \bibnamefont {Casten}},
  \ and\ \bibinfo {author} {\bibfnamefont {E.~A.}\ \bibnamefont {Millman}},\
  }\href {\doibase 10.1103/PhysRevLett.94.092501} {\bibfield  {journal}
  {\bibinfo  {journal} {Phys. Rev. Lett.}\ }\textbf {\bibinfo {volume} {94}},\
  \bibinfo {pages} {092501} (\bibinfo {year} {2005})}\BibitemShut {NoStop}%
\bibitem [{\citenamefont {Cakirli}\ and\ \citenamefont
  {Casten}(2006)}]{Cakirli2006}%
  \BibitemOpen
  \bibfield  {author} {\bibinfo {author} {\bibfnamefont {R.~B.}\ \bibnamefont
  {Cakirli}}\ and\ \bibinfo {author} {\bibfnamefont {R.~F.}\ \bibnamefont
  {Casten}},\ }\href {\doibase 10.1103/PhysRevLett.96.132501} {\bibfield
  {journal} {\bibinfo  {journal} {Phys. Rev. Lett.}\ }\textbf {\bibinfo
  {volume} {96}},\ \bibinfo {pages} {132501} (\bibinfo {year}
  {2006})}\BibitemShut {NoStop}%
\bibitem [{\citenamefont {Stoitsov}\ \emph {et~al.}(2007)\citenamefont
  {Stoitsov}, \citenamefont {Cakirli}, \citenamefont {Casten}, \citenamefont
  {Nazarewicz},\ and\ \citenamefont {Satu\l{}a}}]{Stoitsov2007}%
  \BibitemOpen
  \bibfield  {author} {\bibinfo {author} {\bibfnamefont {M.}~\bibnamefont
  {Stoitsov}}, \bibinfo {author} {\bibfnamefont {R.~B.}\ \bibnamefont
  {Cakirli}}, \bibinfo {author} {\bibfnamefont {R.~F.}\ \bibnamefont {Casten}},
  \bibinfo {author} {\bibfnamefont {W.}~\bibnamefont {Nazarewicz}}, \ and\
  \bibinfo {author} {\bibfnamefont {W.}~\bibnamefont {Satu\l{}a}},\ }\href
  {\doibase 10.1103/PhysRevLett.98.132502} {\bibfield  {journal} {\bibinfo
  {journal} {Phys. Rev. Lett.}\ }\textbf {\bibinfo {volume} {98}},\ \bibinfo
  {pages} {132502} (\bibinfo {year} {2007})}\BibitemShut {NoStop}%
\bibitem [{\citenamefont {Cakirli}\ \emph {et~al.}(2010)\citenamefont
  {Cakirli}, \citenamefont {Blaum},\ and\ \citenamefont
  {Casten}}]{Cakirli2010b}%
  \BibitemOpen
  \bibfield  {author} {\bibinfo {author} {\bibfnamefont {R.~B.}\ \bibnamefont
  {Cakirli}}, \bibinfo {author} {\bibfnamefont {K.}~\bibnamefont {Blaum}}, \
  and\ \bibinfo {author} {\bibfnamefont {R.~F.}\ \bibnamefont {Casten}},\
  }\href {\doibase 10.1103/PhysRevC.82.061304} {\bibfield  {journal} {\bibinfo
  {journal} {Phys. Rev. C}\ }\textbf {\bibinfo {volume} {82}},\ \bibinfo
  {pages} {061304} (\bibinfo {year} {2010})}\BibitemShut {NoStop}%
\bibitem [{\citenamefont {Brenner}\ \emph {et~al.}(2006)\citenamefont
  {Brenner}, \citenamefont {Cakirli},\ and\ \citenamefont
  {Casten}}]{Brenner2006}%
  \BibitemOpen
  \bibfield  {author} {\bibinfo {author} {\bibfnamefont {D.~S.}\ \bibnamefont
  {Brenner}}, \bibinfo {author} {\bibfnamefont {R.~B.}\ \bibnamefont
  {Cakirli}}, \ and\ \bibinfo {author} {\bibfnamefont {R.~F.}\ \bibnamefont
  {Casten}},\ }\href {\doibase 10.1103/PhysRevC.73.034315} {\bibfield
  {journal} {\bibinfo  {journal} {Phys. Rev. C}\ }\textbf {\bibinfo {volume}
  {73}},\ \bibinfo {pages} {034315} (\bibinfo {year} {2006})}\BibitemShut
  {NoStop}%
\bibitem [{\citenamefont {Chen}\ \emph {et~al.}(2009)\citenamefont {Chen},
  \citenamefont {Litvinov}, \citenamefont {Pla\ss{}}, \citenamefont {Beckert},
  \citenamefont {Beller}, \citenamefont {Bosch}, \citenamefont {Boutin},
  \citenamefont {Caceres}, \citenamefont {Cakirli}, \citenamefont {Carroll},
  \citenamefont {Casten}, \citenamefont {Chakrawarthy}, \citenamefont {Cullen},
  \citenamefont {Cullen}, \citenamefont {Franzke}, \citenamefont {Geissel},
  \citenamefont {Gerl}, \citenamefont {G\'orska}, \citenamefont {Jones},
  \citenamefont {Kishada}, \citenamefont {Kn\"obel}, \citenamefont
  {Kozhuharov}, \citenamefont {Litvinov}, \citenamefont {Liu}, \citenamefont
  {Mandal}, \citenamefont {Montes}, \citenamefont {M\"unzenberg}, \citenamefont
  {Nolden}, \citenamefont {Ohtsubo}, \citenamefont {Patyk}, \citenamefont
  {Podoly\'ak}, \citenamefont {Propri}, \citenamefont {Rigby}, \citenamefont
  {Saito}, \citenamefont {Saito}, \citenamefont {Scheidenberger}, \citenamefont
  {Shindo}, \citenamefont {Steck}, \citenamefont {Ugorowski}, \citenamefont
  {Walker}, \citenamefont {Williams}, \citenamefont {Weick}, \citenamefont
  {Winkler}, \citenamefont {Wollersheim},\ and\ \citenamefont
  {Yamaguchi}}]{Chen2009}%
  \BibitemOpen
  \bibfield  {author} {\bibinfo {author} {\bibfnamefont {L.}~\bibnamefont
  {Chen}}, \bibinfo {author} {\bibfnamefont {Y.~A.}\ \bibnamefont {Litvinov}},
  \bibinfo {author} {\bibfnamefont {W.~R.}\ \bibnamefont {Pla\ss{}}}, \bibinfo
  {author} {\bibfnamefont {K.}~\bibnamefont {Beckert}}, \bibinfo {author}
  {\bibfnamefont {P.}~\bibnamefont {Beller}}, \bibinfo {author} {\bibfnamefont
  {F.}~\bibnamefont {Bosch}}, \bibinfo {author} {\bibfnamefont
  {D.}~\bibnamefont {Boutin}}, \bibinfo {author} {\bibfnamefont
  {L.}~\bibnamefont {Caceres}}, \bibinfo {author} {\bibfnamefont {R.~B.}\
  \bibnamefont {Cakirli}}, \bibinfo {author} {\bibfnamefont {J.~J.}\
  \bibnamefont {Carroll}}, \bibinfo {author} {\bibfnamefont {R.~F.}\
  \bibnamefont {Casten}}, \bibinfo {author} {\bibfnamefont {R.~S.}\
  \bibnamefont {Chakrawarthy}}, \bibinfo {author} {\bibfnamefont {D.~M.}\
  \bibnamefont {Cullen}}, \bibinfo {author} {\bibfnamefont {I.~J.}\
  \bibnamefont {Cullen}}, \bibinfo {author} {\bibfnamefont {B.}~\bibnamefont
  {Franzke}}, \bibinfo {author} {\bibfnamefont {H.}~\bibnamefont {Geissel}},
  \bibinfo {author} {\bibfnamefont {J.}~\bibnamefont {Gerl}}, \bibinfo {author}
  {\bibfnamefont {M.}~\bibnamefont {G\'orska}}, \bibinfo {author}
  {\bibfnamefont {G.~A.}\ \bibnamefont {Jones}}, \bibinfo {author}
  {\bibfnamefont {A.}~\bibnamefont {Kishada}}, \bibinfo {author} {\bibfnamefont
  {R.}~\bibnamefont {Kn\"obel}}, \bibinfo {author} {\bibfnamefont
  {C.}~\bibnamefont {Kozhuharov}}, \bibinfo {author} {\bibfnamefont {S.~A.}\
  \bibnamefont {Litvinov}}, \bibinfo {author} {\bibfnamefont {Z.}~\bibnamefont
  {Liu}}, \bibinfo {author} {\bibfnamefont {S.}~\bibnamefont {Mandal}},
  \bibinfo {author} {\bibfnamefont {F.}~\bibnamefont {Montes}}, \bibinfo
  {author} {\bibfnamefont {G.}~\bibnamefont {M\"unzenberg}}, \bibinfo {author}
  {\bibfnamefont {F.}~\bibnamefont {Nolden}}, \bibinfo {author} {\bibfnamefont
  {T.}~\bibnamefont {Ohtsubo}}, \bibinfo {author} {\bibfnamefont
  {Z.}~\bibnamefont {Patyk}}, \bibinfo {author} {\bibfnamefont
  {Z.}~\bibnamefont {Podoly\'ak}}, \bibinfo {author} {\bibfnamefont
  {R.}~\bibnamefont {Propri}}, \bibinfo {author} {\bibfnamefont
  {S.}~\bibnamefont {Rigby}}, \bibinfo {author} {\bibfnamefont
  {N.}~\bibnamefont {Saito}}, \bibinfo {author} {\bibfnamefont
  {T.}~\bibnamefont {Saito}}, \bibinfo {author} {\bibfnamefont
  {C.}~\bibnamefont {Scheidenberger}}, \bibinfo {author} {\bibfnamefont
  {M.}~\bibnamefont {Shindo}}, \bibinfo {author} {\bibfnamefont
  {M.}~\bibnamefont {Steck}}, \bibinfo {author} {\bibfnamefont
  {P.}~\bibnamefont {Ugorowski}}, \bibinfo {author} {\bibfnamefont {P.~M.}\
  \bibnamefont {Walker}}, \bibinfo {author} {\bibfnamefont {S.}~\bibnamefont
  {Williams}}, \bibinfo {author} {\bibfnamefont {H.}~\bibnamefont {Weick}},
  \bibinfo {author} {\bibfnamefont {M.}~\bibnamefont {Winkler}}, \bibinfo
  {author} {\bibfnamefont {H.-J.}\ \bibnamefont {Wollersheim}}, \ and\ \bibinfo
  {author} {\bibfnamefont {T.}~\bibnamefont {Yamaguchi}},\ }\href {\doibase
  10.1103/PhysRevLett.102.122503} {\bibfield  {journal} {\bibinfo  {journal}
  {Phys. Rev. Lett.}\ }\textbf {\bibinfo {volume} {102}},\ \bibinfo {pages}
  {122503} (\bibinfo {year} {2009})}\BibitemShut {NoStop}%
\bibitem [{\citenamefont {Zhang}\ \emph {et~al.}(2018)\citenamefont {Zhang},
  \citenamefont {Zhang}, \citenamefont {Zhou}, \citenamefont {Wang},
  \citenamefont {Litvinov}, \citenamefont {Xu}, \citenamefont {Xu},
  \citenamefont {Shuai}, \citenamefont {Lam}, \citenamefont {Chen},
  \citenamefont {Yan}, \citenamefont {Bao}, \citenamefont {Chen}, \citenamefont
  {Chen}, \citenamefont {Fu}, \citenamefont {He}, \citenamefont {Kubono},
  \citenamefont {Liu}, \citenamefont {Mao}, \citenamefont {Ma}, \citenamefont
  {Sun}, \citenamefont {Tu}, \citenamefont {Xing}, \citenamefont {Zeng},
  \citenamefont {Zhou}, \citenamefont {Zhan}, \citenamefont {Litvinov},
  \citenamefont {Blaum}, \citenamefont {Audi}, \citenamefont {Uesaka},
  \citenamefont {Yamaguchi}, \citenamefont {Yamaguchi}, \citenamefont {Ozawa},
  \citenamefont {Sun}, \citenamefont {Sun},\ and\ \citenamefont
  {Xu}}]{Zhang2018}%
  \BibitemOpen
  \bibfield  {author} {\bibinfo {author} {\bibfnamefont {Y.~H.}\ \bibnamefont
  {Zhang}}, \bibinfo {author} {\bibfnamefont {P.}~\bibnamefont {Zhang}},
  \bibinfo {author} {\bibfnamefont {X.~H.}\ \bibnamefont {Zhou}}, \bibinfo
  {author} {\bibfnamefont {M.}~\bibnamefont {Wang}}, \bibinfo {author}
  {\bibfnamefont {Y.~A.}\ \bibnamefont {Litvinov}}, \bibinfo {author}
  {\bibfnamefont {H.~S.}\ \bibnamefont {Xu}}, \bibinfo {author} {\bibfnamefont
  {X.}~\bibnamefont {Xu}}, \bibinfo {author} {\bibfnamefont {P.}~\bibnamefont
  {Shuai}}, \bibinfo {author} {\bibfnamefont {Y.~H.}\ \bibnamefont {Lam}},
  \bibinfo {author} {\bibfnamefont {R.~J.}\ \bibnamefont {Chen}}, \bibinfo
  {author} {\bibfnamefont {X.~L.}\ \bibnamefont {Yan}}, \bibinfo {author}
  {\bibfnamefont {T.}~\bibnamefont {Bao}}, \bibinfo {author} {\bibfnamefont
  {X.~C.}\ \bibnamefont {Chen}}, \bibinfo {author} {\bibfnamefont
  {H.}~\bibnamefont {Chen}}, \bibinfo {author} {\bibfnamefont {C.~Y.}\
  \bibnamefont {Fu}}, \bibinfo {author} {\bibfnamefont {J.~J.}\ \bibnamefont
  {He}}, \bibinfo {author} {\bibfnamefont {S.}~\bibnamefont {Kubono}}, \bibinfo
  {author} {\bibfnamefont {D.~W.}\ \bibnamefont {Liu}}, \bibinfo {author}
  {\bibfnamefont {R.~S.}\ \bibnamefont {Mao}}, \bibinfo {author} {\bibfnamefont
  {X.~W.}\ \bibnamefont {Ma}}, \bibinfo {author} {\bibfnamefont {M.~Z.}\
  \bibnamefont {Sun}}, \bibinfo {author} {\bibfnamefont {X.~L.}\ \bibnamefont
  {Tu}}, \bibinfo {author} {\bibfnamefont {Y.~M.}\ \bibnamefont {Xing}},
  \bibinfo {author} {\bibfnamefont {Q.}~\bibnamefont {Zeng}}, \bibinfo {author}
  {\bibfnamefont {X.}~\bibnamefont {Zhou}}, \bibinfo {author} {\bibfnamefont
  {W.~L.}\ \bibnamefont {Zhan}}, \bibinfo {author} {\bibfnamefont
  {S.}~\bibnamefont {Litvinov}}, \bibinfo {author} {\bibfnamefont
  {K.}~\bibnamefont {Blaum}}, \bibinfo {author} {\bibfnamefont
  {G.}~\bibnamefont {Audi}}, \bibinfo {author} {\bibfnamefont {T.}~\bibnamefont
  {Uesaka}}, \bibinfo {author} {\bibfnamefont {Y.}~\bibnamefont {Yamaguchi}},
  \bibinfo {author} {\bibfnamefont {T.}~\bibnamefont {Yamaguchi}}, \bibinfo
  {author} {\bibfnamefont {A.}~\bibnamefont {Ozawa}}, \bibinfo {author}
  {\bibfnamefont {B.~H.}\ \bibnamefont {Sun}}, \bibinfo {author} {\bibfnamefont
  {Y.}~\bibnamefont {Sun}}, \ and\ \bibinfo {author} {\bibfnamefont {F.~R.}\
  \bibnamefont {Xu}},\ }\href {\doibase 10.1103/PhysRevC.98.014319} {\bibfield
  {journal} {\bibinfo  {journal} {Phys. Rev. C}\ }\textbf {\bibinfo {volume}
  {98}},\ \bibinfo {pages} {014319} (\bibinfo {year} {2018})}\BibitemShut
  {NoStop}%
\bibitem [{\citenamefont {Schury}\ \emph {et~al.}(2007)\citenamefont {Schury},
  \citenamefont {Bachelet}, \citenamefont {Block}, \citenamefont {Bollen},
  \citenamefont {Davies}, \citenamefont {Facina}, \citenamefont {Folden~III},
  \citenamefont {Gu\'enaut}, \citenamefont {Huikari}, \citenamefont {Kwan},
  \citenamefont {Kwiatkowski}, \citenamefont {Morrissey}, \citenamefont
  {Ringle}, \citenamefont {Pang}, \citenamefont {Prinke}, \citenamefont
  {Savory}, \citenamefont {Schatz}, \citenamefont {Schwarz}, \citenamefont
  {Sumithrarachchi},\ and\ \citenamefont {Sun}}]{Schury2007}%
  \BibitemOpen
  \bibfield  {author} {\bibinfo {author} {\bibfnamefont {P.}~\bibnamefont
  {Schury}}, \bibinfo {author} {\bibfnamefont {C.}~\bibnamefont {Bachelet}},
  \bibinfo {author} {\bibfnamefont {M.}~\bibnamefont {Block}}, \bibinfo
  {author} {\bibfnamefont {G.}~\bibnamefont {Bollen}}, \bibinfo {author}
  {\bibfnamefont {D.~A.}\ \bibnamefont {Davies}}, \bibinfo {author}
  {\bibfnamefont {M.}~\bibnamefont {Facina}}, \bibinfo {author} {\bibfnamefont
  {C.~M.}\ \bibnamefont {Folden~III}}, \bibinfo {author} {\bibfnamefont
  {C.}~\bibnamefont {Gu\'enaut}}, \bibinfo {author} {\bibfnamefont
  {J.}~\bibnamefont {Huikari}}, \bibinfo {author} {\bibfnamefont
  {E.}~\bibnamefont {Kwan}}, \bibinfo {author} {\bibfnamefont {A.}~\bibnamefont
  {Kwiatkowski}}, \bibinfo {author} {\bibfnamefont {D.~J.}\ \bibnamefont
  {Morrissey}}, \bibinfo {author} {\bibfnamefont {R.}~\bibnamefont {Ringle}},
  \bibinfo {author} {\bibfnamefont {G.~K.}\ \bibnamefont {Pang}}, \bibinfo
  {author} {\bibfnamefont {A.}~\bibnamefont {Prinke}}, \bibinfo {author}
  {\bibfnamefont {J.}~\bibnamefont {Savory}}, \bibinfo {author} {\bibfnamefont
  {H.}~\bibnamefont {Schatz}}, \bibinfo {author} {\bibfnamefont
  {S.}~\bibnamefont {Schwarz}}, \bibinfo {author} {\bibfnamefont {C.~S.}\
  \bibnamefont {Sumithrarachchi}}, \ and\ \bibinfo {author} {\bibfnamefont
  {T.}~\bibnamefont {Sun}},\ }\href {\doibase 10.1103/PhysRevC.75.055801}
  {\bibfield  {journal} {\bibinfo  {journal} {Phys. Rev. C}\ }\textbf {\bibinfo
  {volume} {75}},\ \bibinfo {pages} {055801} (\bibinfo {year}
  {2007})}\BibitemShut {NoStop}%
\bibitem [{\citenamefont {Tu}\ \emph {et~al.}(2011)\citenamefont {Tu},
  \citenamefont {Xu}, \citenamefont {Wang}, \citenamefont {Zhang},
  \citenamefont {Litvinov}, \citenamefont {Sun}, \citenamefont {Schatz},
  \citenamefont {Zhou}, \citenamefont {Yuan}, \citenamefont {Xia},
  \citenamefont {Audi}, \citenamefont {Blaum}, \citenamefont {Du},
  \citenamefont {Geng}, \citenamefont {Hu}, \citenamefont {Huang},
  \citenamefont {Jin}, \citenamefont {Liu}, \citenamefont {Liu}, \citenamefont
  {Ma}, \citenamefont {Mao}, \citenamefont {Mei}, \citenamefont {Shuai},
  \citenamefont {Sun}, \citenamefont {Suzuki}, \citenamefont {Tang},
  \citenamefont {Wang}, \citenamefont {Wang}, \citenamefont {Xiao},
  \citenamefont {Xu}, \citenamefont {Yamaguchi}, \citenamefont {Yamaguchi},
  \citenamefont {Yan}, \citenamefont {Yang}, \citenamefont {Ye}, \citenamefont
  {Zang}, \citenamefont {Zhao}, \citenamefont {Zhao}, \citenamefont {Zhang},\
  and\ \citenamefont {Zhan}}]{Tu2011}%
  \BibitemOpen
  \bibfield  {author} {\bibinfo {author} {\bibfnamefont {X.~L.}\ \bibnamefont
  {Tu}}, \bibinfo {author} {\bibfnamefont {H.~S.}\ \bibnamefont {Xu}}, \bibinfo
  {author} {\bibfnamefont {M.}~\bibnamefont {Wang}}, \bibinfo {author}
  {\bibfnamefont {Y.~H.}\ \bibnamefont {Zhang}}, \bibinfo {author}
  {\bibfnamefont {Y.~A.}\ \bibnamefont {Litvinov}}, \bibinfo {author}
  {\bibfnamefont {Y.}~\bibnamefont {Sun}}, \bibinfo {author} {\bibfnamefont
  {H.}~\bibnamefont {Schatz}}, \bibinfo {author} {\bibfnamefont {X.~H.}\
  \bibnamefont {Zhou}}, \bibinfo {author} {\bibfnamefont {Y.~J.}\ \bibnamefont
  {Yuan}}, \bibinfo {author} {\bibfnamefont {J.~W.}\ \bibnamefont {Xia}},
  \bibinfo {author} {\bibfnamefont {G.}~\bibnamefont {Audi}}, \bibinfo {author}
  {\bibfnamefont {K.}~\bibnamefont {Blaum}}, \bibinfo {author} {\bibfnamefont
  {C.~M.}\ \bibnamefont {Du}}, \bibinfo {author} {\bibfnamefont
  {P.}~\bibnamefont {Geng}}, \bibinfo {author} {\bibfnamefont {Z.~G.}\
  \bibnamefont {Hu}}, \bibinfo {author} {\bibfnamefont {W.~X.}\ \bibnamefont
  {Huang}}, \bibinfo {author} {\bibfnamefont {S.~L.}\ \bibnamefont {Jin}},
  \bibinfo {author} {\bibfnamefont {L.~X.}\ \bibnamefont {Liu}}, \bibinfo
  {author} {\bibfnamefont {Y.}~\bibnamefont {Liu}}, \bibinfo {author}
  {\bibfnamefont {X.}~\bibnamefont {Ma}}, \bibinfo {author} {\bibfnamefont
  {R.~S.}\ \bibnamefont {Mao}}, \bibinfo {author} {\bibfnamefont
  {B.}~\bibnamefont {Mei}}, \bibinfo {author} {\bibfnamefont {P.}~\bibnamefont
  {Shuai}}, \bibinfo {author} {\bibfnamefont {Z.~Y.}\ \bibnamefont {Sun}},
  \bibinfo {author} {\bibfnamefont {H.}~\bibnamefont {Suzuki}}, \bibinfo
  {author} {\bibfnamefont {S.~W.}\ \bibnamefont {Tang}}, \bibinfo {author}
  {\bibfnamefont {J.~S.}\ \bibnamefont {Wang}}, \bibinfo {author}
  {\bibfnamefont {S.~T.}\ \bibnamefont {Wang}}, \bibinfo {author}
  {\bibfnamefont {G.~Q.}\ \bibnamefont {Xiao}}, \bibinfo {author}
  {\bibfnamefont {X.}~\bibnamefont {Xu}}, \bibinfo {author} {\bibfnamefont
  {T.}~\bibnamefont {Yamaguchi}}, \bibinfo {author} {\bibfnamefont
  {Y.}~\bibnamefont {Yamaguchi}}, \bibinfo {author} {\bibfnamefont {X.~L.}\
  \bibnamefont {Yan}}, \bibinfo {author} {\bibfnamefont {J.~C.}\ \bibnamefont
  {Yang}}, \bibinfo {author} {\bibfnamefont {R.~P.}\ \bibnamefont {Ye}},
  \bibinfo {author} {\bibfnamefont {Y.~D.}\ \bibnamefont {Zang}}, \bibinfo
  {author} {\bibfnamefont {H.~W.}\ \bibnamefont {Zhao}}, \bibinfo {author}
  {\bibfnamefont {T.~C.}\ \bibnamefont {Zhao}}, \bibinfo {author}
  {\bibfnamefont {X.~Y.}\ \bibnamefont {Zhang}}, \ and\ \bibinfo {author}
  {\bibfnamefont {W.~L.}\ \bibnamefont {Zhan}},\ }\href {\doibase
  10.1103/PhysRevLett.106.112501} {\bibfield  {journal} {\bibinfo  {journal}
  {Phys. Rev. Lett.}\ }\textbf {\bibinfo {volume} {106}},\ \bibinfo {pages}
  {112501} (\bibinfo {year} {2011})}\BibitemShut {NoStop}%
\bibitem [{\citenamefont {Hoff}\ \emph {et~al.}(2020)\citenamefont {Hoff},
  \citenamefont {Rogers}, \citenamefont {Meisel}, \citenamefont {Bender},
  \citenamefont {Brandenburg}, \citenamefont {Childers}, \citenamefont {Clark},
  \citenamefont {Dombos}, \citenamefont {Doucet}, \citenamefont {Jin},
  \citenamefont {Lewis}, \citenamefont {Liddick}, \citenamefont {Lister},
  \citenamefont {Morse}, \citenamefont {Schatz}, \citenamefont {Schmidt},
  \citenamefont {Soltesz}, \citenamefont {Subedi}, \citenamefont {Wang},\ and\
  \citenamefont {Waniganeththi}}]{Hoff2020}%
  \BibitemOpen
  \bibfield  {author} {\bibinfo {author} {\bibfnamefont {D.~E.~M.}\
  \bibnamefont {Hoff}}, \bibinfo {author} {\bibfnamefont {A.~M.}\ \bibnamefont
  {Rogers}}, \bibinfo {author} {\bibfnamefont {Z.}~\bibnamefont {Meisel}},
  \bibinfo {author} {\bibfnamefont {P.~C.}\ \bibnamefont {Bender}}, \bibinfo
  {author} {\bibfnamefont {K.}~\bibnamefont {Brandenburg}}, \bibinfo {author}
  {\bibfnamefont {K.}~\bibnamefont {Childers}}, \bibinfo {author}
  {\bibfnamefont {J.~A.}\ \bibnamefont {Clark}}, \bibinfo {author}
  {\bibfnamefont {A.~C.}\ \bibnamefont {Dombos}}, \bibinfo {author}
  {\bibfnamefont {E.~R.}\ \bibnamefont {Doucet}}, \bibinfo {author}
  {\bibfnamefont {S.}~\bibnamefont {Jin}}, \bibinfo {author} {\bibfnamefont
  {R.}~\bibnamefont {Lewis}}, \bibinfo {author} {\bibfnamefont {S.~N.}\
  \bibnamefont {Liddick}}, \bibinfo {author} {\bibfnamefont {C.~J.}\
  \bibnamefont {Lister}}, \bibinfo {author} {\bibfnamefont {C.}~\bibnamefont
  {Morse}}, \bibinfo {author} {\bibfnamefont {H.}~\bibnamefont {Schatz}},
  \bibinfo {author} {\bibfnamefont {K.}~\bibnamefont {Schmidt}}, \bibinfo
  {author} {\bibfnamefont {D.}~\bibnamefont {Soltesz}}, \bibinfo {author}
  {\bibfnamefont {S.~K.}\ \bibnamefont {Subedi}}, \bibinfo {author}
  {\bibfnamefont {S.~M.}\ \bibnamefont {Wang}}, \ and\ \bibinfo {author}
  {\bibfnamefont {S.}~\bibnamefont {Waniganeththi}},\ }\href {\doibase
  10.1103/PhysRevC.102.045810} {\bibfield  {journal} {\bibinfo  {journal}
  {Phys. Rev. C}\ }\textbf {\bibinfo {volume} {102}},\ \bibinfo {pages}
  {045810} (\bibinfo {year} {2020})}\BibitemShut {NoStop}%
\bibitem [{\citenamefont {Wang}\ \emph {et~al.}(2017)\citenamefont {Wang},
  \citenamefont {Audi}, \citenamefont {Kondev}, \citenamefont {Huang},
  \citenamefont {Naimi},\ and\ \citenamefont {Xu}}]{Wang2017}%
  \BibitemOpen
  \bibfield  {author} {\bibinfo {author} {\bibfnamefont {M.}~\bibnamefont
  {Wang}}, \bibinfo {author} {\bibfnamefont {G.}~\bibnamefont {Audi}}, \bibinfo
  {author} {\bibfnamefont {F.~G.}\ \bibnamefont {Kondev}}, \bibinfo {author}
  {\bibfnamefont {W.~J.}\ \bibnamefont {Huang}}, \bibinfo {author}
  {\bibfnamefont {S.}~\bibnamefont {Naimi}}, \ and\ \bibinfo {author}
  {\bibfnamefont {X.}~\bibnamefont {Xu}},\ }\href@noop {} {\bibfield  {journal}
  {\bibinfo  {journal} {{Chinese Phys. C}}\ }\textbf {\bibinfo {volume} {{41}}}
  (\bibinfo {year} {2017})}\BibitemShut {NoStop}%
\bibitem [{\citenamefont {Wollnik}\ and\ \citenamefont
  {Przewloka}(1990)}]{Wollnik1990}%
  \BibitemOpen
  \bibfield  {author} {\bibinfo {author} {\bibfnamefont {H.}~\bibnamefont
  {Wollnik}}\ and\ \bibinfo {author} {\bibfnamefont {M.}~\bibnamefont
  {Przewloka}},\ }\href@noop {} {\bibfield  {journal} {\bibinfo  {journal}
  {Int. J. Mass Spectrom. Ion Processes}\ }\textbf {\bibinfo {volume} {96}},\
  \bibinfo {pages} {267 } (\bibinfo {year} {1990})}\BibitemShut {NoStop}%
\bibitem [{\citenamefont {Pla\ss{}}\ \emph
  {et~al.}(2013{\natexlab{a}})\citenamefont {Pla\ss{}}, \citenamefont
  {Dickel},\ and\ \citenamefont {Scheidenberger}}]{Plass2013b}%
  \BibitemOpen
  \bibfield  {author} {\bibinfo {author} {\bibfnamefont {W.~R.}\ \bibnamefont
  {Pla\ss{}}}, \bibinfo {author} {\bibfnamefont {T.}~\bibnamefont {Dickel}}, \
  and\ \bibinfo {author} {\bibfnamefont {C.}~\bibnamefont {Scheidenberger}},\
  }\href@noop {} {\bibfield  {journal} {\bibinfo  {journal} {Int. J. of Mass
  Spectrom.}\ }\textbf {\bibinfo {volume} {349 - 350}},\ \bibinfo {pages} {134
  } (\bibinfo {year} {2013}{\natexlab{a}})}\BibitemShut {NoStop}%
\bibitem [{\citenamefont {Hornung}\ \emph {et~al.}(2020)\citenamefont
  {Hornung}, \citenamefont {Amanbayev}, \citenamefont {Dedes}, \citenamefont
  {Kripko-Koncz}, \citenamefont {Miskun}, \citenamefont {Shimizu},
  \citenamefont {{Ayet San Andrés}}, \citenamefont {Bergmann}, \citenamefont
  {Dickel}, \citenamefont {Dudek}, \citenamefont {Ebert}, \citenamefont
  {Geissel}, \citenamefont {Górska}, \citenamefont {Grawe}, \citenamefont
  {Greiner}, \citenamefont {Haettner}, \citenamefont {Otsuka}, \citenamefont
  {Plaß}, \citenamefont {Purushothaman}, \citenamefont {Rink}, \citenamefont
  {Scheidenberger}, \citenamefont {Weick}, \citenamefont {Bagchi},
  \citenamefont {Blazhev}, \citenamefont {Charviakova}, \citenamefont {Curien},
  \citenamefont {Finlay}, \citenamefont {Kaur}, \citenamefont {Lippert},
  \citenamefont {Otto}, \citenamefont {Patyk}, \citenamefont {Pietri},
  \citenamefont {Tanaka}, \citenamefont {Tsunoda},\ and\ \citenamefont
  {Winfield}}]{Hornung2020}%
  \BibitemOpen
  \bibfield  {author} {\bibinfo {author} {\bibfnamefont {C.}~\bibnamefont
  {Hornung}}, \bibinfo {author} {\bibfnamefont {D.}~\bibnamefont {Amanbayev}},
  \bibinfo {author} {\bibfnamefont {I.}~\bibnamefont {Dedes}}, \bibinfo
  {author} {\bibfnamefont {G.}~\bibnamefont {Kripko-Koncz}}, \bibinfo {author}
  {\bibfnamefont {I.}~\bibnamefont {Miskun}}, \bibinfo {author} {\bibfnamefont
  {N.}~\bibnamefont {Shimizu}}, \bibinfo {author} {\bibfnamefont
  {S.}~\bibnamefont {{Ayet San Andrés}}}, \bibinfo {author} {\bibfnamefont
  {J.}~\bibnamefont {Bergmann}}, \bibinfo {author} {\bibfnamefont
  {T.}~\bibnamefont {Dickel}}, \bibinfo {author} {\bibfnamefont
  {J.}~\bibnamefont {Dudek}}, \bibinfo {author} {\bibfnamefont
  {J.}~\bibnamefont {Ebert}}, \bibinfo {author} {\bibfnamefont
  {H.}~\bibnamefont {Geissel}}, \bibinfo {author} {\bibfnamefont
  {M.}~\bibnamefont {Górska}}, \bibinfo {author} {\bibfnamefont
  {H.}~\bibnamefont {Grawe}}, \bibinfo {author} {\bibfnamefont
  {F.}~\bibnamefont {Greiner}}, \bibinfo {author} {\bibfnamefont
  {E.}~\bibnamefont {Haettner}}, \bibinfo {author} {\bibfnamefont
  {T.}~\bibnamefont {Otsuka}}, \bibinfo {author} {\bibfnamefont {W.~R.}\
  \bibnamefont {Plaß}}, \bibinfo {author} {\bibfnamefont {S.}~\bibnamefont
  {Purushothaman}}, \bibinfo {author} {\bibfnamefont {A.-K.}\ \bibnamefont
  {Rink}}, \bibinfo {author} {\bibfnamefont {C.}~\bibnamefont
  {Scheidenberger}}, \bibinfo {author} {\bibfnamefont {H.}~\bibnamefont
  {Weick}}, \bibinfo {author} {\bibfnamefont {S.}~\bibnamefont {Bagchi}},
  \bibinfo {author} {\bibfnamefont {A.}~\bibnamefont {Blazhev}}, \bibinfo
  {author} {\bibfnamefont {O.}~\bibnamefont {Charviakova}}, \bibinfo {author}
  {\bibfnamefont {D.}~\bibnamefont {Curien}}, \bibinfo {author} {\bibfnamefont
  {A.}~\bibnamefont {Finlay}}, \bibinfo {author} {\bibfnamefont
  {S.}~\bibnamefont {Kaur}}, \bibinfo {author} {\bibfnamefont {W.}~\bibnamefont
  {Lippert}}, \bibinfo {author} {\bibfnamefont {J.-H.}\ \bibnamefont {Otto}},
  \bibinfo {author} {\bibfnamefont {Z.}~\bibnamefont {Patyk}}, \bibinfo
  {author} {\bibfnamefont {S.}~\bibnamefont {Pietri}}, \bibinfo {author}
  {\bibfnamefont {Y.~K.}\ \bibnamefont {Tanaka}}, \bibinfo {author}
  {\bibfnamefont {Y.}~\bibnamefont {Tsunoda}}, \ and\ \bibinfo {author}
  {\bibfnamefont {J.~S.}\ \bibnamefont {Winfield}},\ }\href {\doibase
  https://doi.org/10.1016/j.physletb.2020.135200} {\bibfield  {journal}
  {\bibinfo  {journal} {Phys. Lett. B}\ }\textbf {\bibinfo {volume} {802}},\
  \bibinfo {pages} {135200} (\bibinfo {year} {2020})}\BibitemShut {NoStop}%
\bibitem [{\citenamefont {Dickel}\ \emph
  {et~al.}(2015{\natexlab{a}})\citenamefont {Dickel}, \citenamefont {Pla\ss{}},
  \citenamefont {Ayet San~Andr\'es}, \citenamefont {Ebert}, \citenamefont
  {Geissel}, \citenamefont {Haettner}, \citenamefont {Hornung}, \citenamefont
  {Miskun}, \citenamefont {Pietri}, \citenamefont {Purushothaman},
  \citenamefont {Reiter}, \citenamefont {Rink}, \citenamefont {Scheidenberger},
  \citenamefont {Weick}, \citenamefont {Dendooven}, \citenamefont {Diwisch},
  \citenamefont {Greiner}, \citenamefont {Heisse}, \citenamefont {Kn\"obel},
  \citenamefont {Lippert}, \citenamefont {Moore}, \citenamefont {Pohjalainen},
  \citenamefont {Prochazka}, \citenamefont {Ranjan}, \citenamefont {Takechi},
  \citenamefont {Winfield},\ and\ \citenamefont {Xu}}]{Dickel2015}%
  \BibitemOpen
  \bibfield  {author} {\bibinfo {author} {\bibfnamefont {T.}~\bibnamefont
  {Dickel}}, \bibinfo {author} {\bibfnamefont {W.~R.}\ \bibnamefont
  {Pla\ss{}}}, \bibinfo {author} {\bibfnamefont {S.}~\bibnamefont {Ayet
  San~Andr\'es}}, \bibinfo {author} {\bibfnamefont {J.}~\bibnamefont {Ebert}},
  \bibinfo {author} {\bibfnamefont {H.}~\bibnamefont {Geissel}}, \bibinfo
  {author} {\bibfnamefont {E.}~\bibnamefont {Haettner}}, \bibinfo {author}
  {\bibfnamefont {C.}~\bibnamefont {Hornung}}, \bibinfo {author} {\bibfnamefont
  {I.}~\bibnamefont {Miskun}}, \bibinfo {author} {\bibfnamefont
  {S.}~\bibnamefont {Pietri}}, \bibinfo {author} {\bibfnamefont
  {S.}~\bibnamefont {Purushothaman}}, \bibinfo {author} {\bibfnamefont {M.~P.}\
  \bibnamefont {Reiter}}, \bibinfo {author} {\bibfnamefont {A.~K.}\
  \bibnamefont {Rink}}, \bibinfo {author} {\bibfnamefont {C.}~\bibnamefont
  {Scheidenberger}}, \bibinfo {author} {\bibfnamefont {H.}~\bibnamefont
  {Weick}}, \bibinfo {author} {\bibfnamefont {P.}~\bibnamefont {Dendooven}},
  \bibinfo {author} {\bibfnamefont {M.}~\bibnamefont {Diwisch}}, \bibinfo
  {author} {\bibfnamefont {F.}~\bibnamefont {Greiner}}, \bibinfo {author}
  {\bibfnamefont {F.}~\bibnamefont {Heisse}}, \bibinfo {author} {\bibfnamefont
  {R.}~\bibnamefont {Kn\"obel}}, \bibinfo {author} {\bibfnamefont
  {W.}~\bibnamefont {Lippert}}, \bibinfo {author} {\bibfnamefont {I.~D.}\
  \bibnamefont {Moore}}, \bibinfo {author} {\bibfnamefont {I.}~\bibnamefont
  {Pohjalainen}}, \bibinfo {author} {\bibfnamefont {A.}~\bibnamefont
  {Prochazka}}, \bibinfo {author} {\bibfnamefont {M.}~\bibnamefont {Ranjan}},
  \bibinfo {author} {\bibfnamefont {M.}~\bibnamefont {Takechi}}, \bibinfo
  {author} {\bibfnamefont {J.~S.}\ \bibnamefont {Winfield}}, \ and\ \bibinfo
  {author} {\bibfnamefont {X.}~\bibnamefont {Xu}},\ }\href@noop {} {\bibfield
  {journal} {\bibinfo  {journal} {{Phys. Lett. B}}\ }\textbf {\bibinfo {volume}
  {{744}}},\ \bibinfo {pages} {137} (\bibinfo {year}
  {{2015}}{\natexlab{a}})}\BibitemShut {NoStop}%
\bibitem [{\citenamefont {Ayet San~Andr\'es}\ \emph {et~al.}(2019)\citenamefont
  {Ayet San~Andr\'es}, \citenamefont {Hornung}, \citenamefont {Ebert},
  \citenamefont {Pla\ss{}}, \citenamefont {Dickel}, \citenamefont {Geissel},
  \citenamefont {Scheidenberger}, \citenamefont {Bergmann}, \citenamefont
  {Greiner}, \citenamefont {Haettner}, \citenamefont {Jesch}, \citenamefont
  {Lippert}, \citenamefont {Mardor}, \citenamefont {Miskun}, \citenamefont
  {Patyk}, \citenamefont {Pietri}, \citenamefont {Pihktelev}, \citenamefont
  {Purushothaman}, \citenamefont {Reiter}, \citenamefont {Rink}, \citenamefont
  {Weick}, \citenamefont {Yavor}, \citenamefont {Bagchi}, \citenamefont
  {Charviakova}, \citenamefont {Constantin}, \citenamefont {Diwisch},
  \citenamefont {Finlay}, \citenamefont {Kaur}, \citenamefont {Kn\"obel},
  \citenamefont {Lang}, \citenamefont {Mei}, \citenamefont {Moore},
  \citenamefont {Otto}, \citenamefont {Pohjalainen}, \citenamefont {Prochazka},
  \citenamefont {Rappold}, \citenamefont {Takechi}, \citenamefont {Tanaka},
  \citenamefont {Winfield},\ and\ \citenamefont {Xu}}]{Ayet2019}%
  \BibitemOpen
  \bibfield  {author} {\bibinfo {author} {\bibfnamefont {S.}~\bibnamefont {Ayet
  San~Andr\'es}}, \bibinfo {author} {\bibfnamefont {C.}~\bibnamefont
  {Hornung}}, \bibinfo {author} {\bibfnamefont {J.}~\bibnamefont {Ebert}},
  \bibinfo {author} {\bibfnamefont {W.~R.}\ \bibnamefont {Pla\ss{}}}, \bibinfo
  {author} {\bibfnamefont {T.}~\bibnamefont {Dickel}}, \bibinfo {author}
  {\bibfnamefont {H.}~\bibnamefont {Geissel}}, \bibinfo {author} {\bibfnamefont
  {C.}~\bibnamefont {Scheidenberger}}, \bibinfo {author} {\bibfnamefont
  {J.}~\bibnamefont {Bergmann}}, \bibinfo {author} {\bibfnamefont
  {F.}~\bibnamefont {Greiner}}, \bibinfo {author} {\bibfnamefont
  {E.}~\bibnamefont {Haettner}}, \bibinfo {author} {\bibfnamefont
  {C.}~\bibnamefont {Jesch}}, \bibinfo {author} {\bibfnamefont
  {W.}~\bibnamefont {Lippert}}, \bibinfo {author} {\bibfnamefont
  {I.}~\bibnamefont {Mardor}}, \bibinfo {author} {\bibfnamefont
  {I.}~\bibnamefont {Miskun}}, \bibinfo {author} {\bibfnamefont
  {Z.}~\bibnamefont {Patyk}}, \bibinfo {author} {\bibfnamefont
  {S.}~\bibnamefont {Pietri}}, \bibinfo {author} {\bibfnamefont
  {A.}~\bibnamefont {Pihktelev}}, \bibinfo {author} {\bibfnamefont
  {S.}~\bibnamefont {Purushothaman}}, \bibinfo {author} {\bibfnamefont {M.~P.}\
  \bibnamefont {Reiter}}, \bibinfo {author} {\bibfnamefont {A.-K.}\
  \bibnamefont {Rink}}, \bibinfo {author} {\bibfnamefont {H.}~\bibnamefont
  {Weick}}, \bibinfo {author} {\bibfnamefont {M.~I.}\ \bibnamefont {Yavor}},
  \bibinfo {author} {\bibfnamefont {S.}~\bibnamefont {Bagchi}}, \bibinfo
  {author} {\bibfnamefont {V.}~\bibnamefont {Charviakova}}, \bibinfo {author}
  {\bibfnamefont {P.}~\bibnamefont {Constantin}}, \bibinfo {author}
  {\bibfnamefont {M.}~\bibnamefont {Diwisch}}, \bibinfo {author} {\bibfnamefont
  {A.}~\bibnamefont {Finlay}}, \bibinfo {author} {\bibfnamefont
  {S.}~\bibnamefont {Kaur}}, \bibinfo {author} {\bibfnamefont {R.}~\bibnamefont
  {Kn\"obel}}, \bibinfo {author} {\bibfnamefont {J.}~\bibnamefont {Lang}},
  \bibinfo {author} {\bibfnamefont {B.}~\bibnamefont {Mei}}, \bibinfo {author}
  {\bibfnamefont {I.~D.}\ \bibnamefont {Moore}}, \bibinfo {author}
  {\bibfnamefont {J.-H.}\ \bibnamefont {Otto}}, \bibinfo {author}
  {\bibfnamefont {I.}~\bibnamefont {Pohjalainen}}, \bibinfo {author}
  {\bibfnamefont {A.}~\bibnamefont {Prochazka}}, \bibinfo {author}
  {\bibfnamefont {C.}~\bibnamefont {Rappold}}, \bibinfo {author} {\bibfnamefont
  {M.}~\bibnamefont {Takechi}}, \bibinfo {author} {\bibfnamefont {Y.~K.}\
  \bibnamefont {Tanaka}}, \bibinfo {author} {\bibfnamefont {J.~S.}\
  \bibnamefont {Winfield}}, \ and\ \bibinfo {author} {\bibfnamefont
  {X.}~\bibnamefont {Xu}},\ }\href {\doibase 10.1103/PhysRevC.99.064313}
  {\bibfield  {journal} {\bibinfo  {journal} {Phys. Rev. C}\ }\textbf {\bibinfo
  {volume} {99}},\ \bibinfo {pages} {064313} (\bibinfo {year}
  {2019})}\BibitemShut {NoStop}%
\bibitem [{\citenamefont {Geissel}\ \emph {et~al.}(1992)\citenamefont
  {Geissel}, \citenamefont {Armbruster}, \citenamefont {Behr}, \citenamefont
  {Br\"unle}, \citenamefont {Burkard}, \citenamefont {Chen}, \citenamefont
  {Folger}, \citenamefont {Franczak}, \citenamefont {Keller}, \citenamefont
  {Klepper}, \citenamefont {Langenbeck}, \citenamefont {Nickel}, \citenamefont
  {Pfeng}, \citenamefont {Pf\"utzner}, \citenamefont {Roeckl}, \citenamefont
  {Rykaczewski}, \citenamefont {Schall}, \citenamefont {Schardt}, \citenamefont
  {Scheidenberger}, \citenamefont {Schmidt}, \citenamefont {Schr\"oter},
  \citenamefont {Schwab}, \citenamefont {S\"ummerer}, \citenamefont {Weber},
  \citenamefont {M\"unzenberg}, \citenamefont {Brohm}, \citenamefont {Clerc},
  \citenamefont {Fauerbach}, \citenamefont {Gaimard}, \citenamefont {Grewe},
  \citenamefont {Hanelt}, \citenamefont {Kn\"odler}, \citenamefont {Steiner},
  \citenamefont {Voss}, \citenamefont {Weckenmann}, \citenamefont {Ziegler},
  \citenamefont {Magel}, \citenamefont {Wollnik}, \citenamefont {Dufour},
  \citenamefont {Fujita}, \citenamefont {Vieira},\ and\ \citenamefont
  {Sherrill}}]{Geissel1992b}%
  \BibitemOpen
  \bibfield  {author} {\bibinfo {author} {\bibfnamefont {H.}~\bibnamefont
  {Geissel}}, \bibinfo {author} {\bibfnamefont {P.}~\bibnamefont {Armbruster}},
  \bibinfo {author} {\bibfnamefont {K.}~\bibnamefont {Behr}}, \bibinfo {author}
  {\bibfnamefont {A.}~\bibnamefont {Br\"unle}}, \bibinfo {author}
  {\bibfnamefont {K.}~\bibnamefont {Burkard}}, \bibinfo {author} {\bibfnamefont
  {M.}~\bibnamefont {Chen}}, \bibinfo {author} {\bibfnamefont {H.}~\bibnamefont
  {Folger}}, \bibinfo {author} {\bibfnamefont {B.}~\bibnamefont {Franczak}},
  \bibinfo {author} {\bibfnamefont {H.}~\bibnamefont {Keller}}, \bibinfo
  {author} {\bibfnamefont {O.}~\bibnamefont {Klepper}}, \bibinfo {author}
  {\bibfnamefont {B.}~\bibnamefont {Langenbeck}}, \bibinfo {author}
  {\bibfnamefont {F.}~\bibnamefont {Nickel}}, \bibinfo {author} {\bibfnamefont
  {E.}~\bibnamefont {Pfeng}}, \bibinfo {author} {\bibfnamefont
  {M.}~\bibnamefont {Pf\"utzner}}, \bibinfo {author} {\bibfnamefont
  {E.}~\bibnamefont {Roeckl}}, \bibinfo {author} {\bibfnamefont
  {K.}~\bibnamefont {Rykaczewski}}, \bibinfo {author} {\bibfnamefont
  {I.}~\bibnamefont {Schall}}, \bibinfo {author} {\bibfnamefont
  {D.}~\bibnamefont {Schardt}}, \bibinfo {author} {\bibfnamefont
  {C.}~\bibnamefont {Scheidenberger}}, \bibinfo {author} {\bibfnamefont
  {K.-H.}\ \bibnamefont {Schmidt}}, \bibinfo {author} {\bibfnamefont
  {A.}~\bibnamefont {Schr\"oter}}, \bibinfo {author} {\bibfnamefont
  {T.}~\bibnamefont {Schwab}}, \bibinfo {author} {\bibfnamefont
  {K.}~\bibnamefont {S\"ummerer}}, \bibinfo {author} {\bibfnamefont
  {M.}~\bibnamefont {Weber}}, \bibinfo {author} {\bibfnamefont
  {G.}~\bibnamefont {M\"unzenberg}}, \bibinfo {author} {\bibfnamefont
  {T.}~\bibnamefont {Brohm}}, \bibinfo {author} {\bibfnamefont {H.-G.}\
  \bibnamefont {Clerc}}, \bibinfo {author} {\bibfnamefont {M.}~\bibnamefont
  {Fauerbach}}, \bibinfo {author} {\bibfnamefont {J.-J.}\ \bibnamefont
  {Gaimard}}, \bibinfo {author} {\bibfnamefont {A.}~\bibnamefont {Grewe}},
  \bibinfo {author} {\bibfnamefont {E.}~\bibnamefont {Hanelt}}, \bibinfo
  {author} {\bibfnamefont {B.}~\bibnamefont {Kn\"odler}}, \bibinfo {author}
  {\bibfnamefont {M.}~\bibnamefont {Steiner}}, \bibinfo {author} {\bibfnamefont
  {B.}~\bibnamefont {Voss}}, \bibinfo {author} {\bibfnamefont {J.}~\bibnamefont
  {Weckenmann}}, \bibinfo {author} {\bibfnamefont {C.}~\bibnamefont {Ziegler}},
  \bibinfo {author} {\bibfnamefont {A.}~\bibnamefont {Magel}}, \bibinfo
  {author} {\bibfnamefont {H.}~\bibnamefont {Wollnik}}, \bibinfo {author}
  {\bibfnamefont {J.}~\bibnamefont {Dufour}}, \bibinfo {author} {\bibfnamefont
  {Y.}~\bibnamefont {Fujita}}, \bibinfo {author} {\bibfnamefont
  {D.}~\bibnamefont {Vieira}}, \ and\ \bibinfo {author} {\bibfnamefont
  {B.}~\bibnamefont {Sherrill}},\ }\href@noop {} {\bibfield  {journal}
  {\bibinfo  {journal} {Nucl. Instrum. Methods B}\ }\textbf {\bibinfo {volume}
  {70}},\ \bibinfo {pages} {286 } (\bibinfo {year} {1992})}\BibitemShut
  {NoStop}%
\bibitem [{\citenamefont {Pla\ss{}}\ \emph
  {et~al.}(2013{\natexlab{b}})\citenamefont {Pla\ss{}}, \citenamefont {Dickel},
  \citenamefont {Purushothaman}, \citenamefont {Dendooven}, \citenamefont
  {Geissel}, \citenamefont {Ebert}, \citenamefont {Haettner}, \citenamefont
  {Jesch}, \citenamefont {Ranjan}, \citenamefont {Reiter}, \citenamefont
  {Weick}, \citenamefont {Amjad}, \citenamefont {Ayet San~Andr\'es},
  \citenamefont {Diwisch}, \citenamefont {Estrade}, \citenamefont {Farinon},
  \citenamefont {Greiner}, \citenamefont {Kalantar-Nayestanaki}, \citenamefont
  {Kn\"obel}, \citenamefont {Kurcewicz}, \citenamefont {Lang}, \citenamefont
  {Moore}, \citenamefont {Mukha}, \citenamefont {Nociforo}, \citenamefont
  {Petrick}, \citenamefont {Pf\"utzner}, \citenamefont {Pietri}, \citenamefont
  {Prochazka}, \citenamefont {Rink}, \citenamefont {Rinta-Antila},
  \citenamefont {Sch\"afer}, \citenamefont {Scheidenberger}, \citenamefont
  {Takechi}, \citenamefont {Tanaka}, \citenamefont {Winfield},\ and\
  \citenamefont {Yavor}}]{Plass2013}%
  \BibitemOpen
  \bibfield  {author} {\bibinfo {author} {\bibfnamefont {W.~R.}\ \bibnamefont
  {Pla\ss{}}}, \bibinfo {author} {\bibfnamefont {T.}~\bibnamefont {Dickel}},
  \bibinfo {author} {\bibfnamefont {S.}~\bibnamefont {Purushothaman}}, \bibinfo
  {author} {\bibfnamefont {P.}~\bibnamefont {Dendooven}}, \bibinfo {author}
  {\bibfnamefont {H.}~\bibnamefont {Geissel}}, \bibinfo {author} {\bibfnamefont
  {J.}~\bibnamefont {Ebert}}, \bibinfo {author} {\bibfnamefont
  {E.}~\bibnamefont {Haettner}}, \bibinfo {author} {\bibfnamefont
  {C.}~\bibnamefont {Jesch}}, \bibinfo {author} {\bibfnamefont
  {M.}~\bibnamefont {Ranjan}}, \bibinfo {author} {\bibfnamefont {M.~P.}\
  \bibnamefont {Reiter}}, \bibinfo {author} {\bibfnamefont {H.}~\bibnamefont
  {Weick}}, \bibinfo {author} {\bibfnamefont {F.}~\bibnamefont {Amjad}},
  \bibinfo {author} {\bibfnamefont {S.}~\bibnamefont {Ayet San~Andr\'es}},
  \bibinfo {author} {\bibfnamefont {M.}~\bibnamefont {Diwisch}}, \bibinfo
  {author} {\bibfnamefont {A.}~\bibnamefont {Estrade}}, \bibinfo {author}
  {\bibfnamefont {F.}~\bibnamefont {Farinon}}, \bibinfo {author} {\bibfnamefont
  {F.}~\bibnamefont {Greiner}}, \bibinfo {author} {\bibfnamefont
  {N.}~\bibnamefont {Kalantar-Nayestanaki}}, \bibinfo {author} {\bibfnamefont
  {R.}~\bibnamefont {Kn\"obel}}, \bibinfo {author} {\bibfnamefont
  {J.}~\bibnamefont {Kurcewicz}}, \bibinfo {author} {\bibfnamefont
  {J.}~\bibnamefont {Lang}}, \bibinfo {author} {\bibfnamefont {I.}~\bibnamefont
  {Moore}}, \bibinfo {author} {\bibfnamefont {I.}~\bibnamefont {Mukha}},
  \bibinfo {author} {\bibfnamefont {C.}~\bibnamefont {Nociforo}}, \bibinfo
  {author} {\bibfnamefont {M.}~\bibnamefont {Petrick}}, \bibinfo {author}
  {\bibfnamefont {M.}~\bibnamefont {Pf\"utzner}}, \bibinfo {author}
  {\bibfnamefont {S.}~\bibnamefont {Pietri}}, \bibinfo {author} {\bibfnamefont
  {A.}~\bibnamefont {Prochazka}}, \bibinfo {author} {\bibfnamefont {A.-K.}\
  \bibnamefont {Rink}}, \bibinfo {author} {\bibfnamefont {S.}~\bibnamefont
  {Rinta-Antila}}, \bibinfo {author} {\bibfnamefont {D.}~\bibnamefont
  {Sch\"afer}}, \bibinfo {author} {\bibfnamefont {C.}~\bibnamefont
  {Scheidenberger}}, \bibinfo {author} {\bibfnamefont {M.}~\bibnamefont
  {Takechi}}, \bibinfo {author} {\bibfnamefont {Y.~K.}\ \bibnamefont {Tanaka}},
  \bibinfo {author} {\bibfnamefont {J.~S.}\ \bibnamefont {Winfield}}, \ and\
  \bibinfo {author} {\bibfnamefont {M.~I.}\ \bibnamefont {Yavor}},\ }\href@noop
  {} {\bibfield  {journal} {\bibinfo  {journal} {Nucl. Instrum. Methods B}\
  }\textbf {\bibinfo {volume} {317}},\ \bibinfo {pages} {457} (\bibinfo {year}
  {2013}{\natexlab{b}})}\BibitemShut {NoStop}%
\bibitem [{\citenamefont {Ranjan}\ \emph {et~al.}(2011)\citenamefont {Ranjan},
  \citenamefont {Purushothaman}, \citenamefont {Dickel}, \citenamefont
  {Geissel}, \citenamefont {Pla\ss{}}, \citenamefont {Sch\"afer}, \citenamefont
  {Scheidenberger}, \citenamefont {de~Walle}, \citenamefont {Weick},\ and\
  \citenamefont {Dendooven}}]{Ranjan2011}%
  \BibitemOpen
  \bibfield  {author} {\bibinfo {author} {\bibfnamefont {M.}~\bibnamefont
  {Ranjan}}, \bibinfo {author} {\bibfnamefont {S.}~\bibnamefont
  {Purushothaman}}, \bibinfo {author} {\bibfnamefont {T.}~\bibnamefont
  {Dickel}}, \bibinfo {author} {\bibfnamefont {H.}~\bibnamefont {Geissel}},
  \bibinfo {author} {\bibfnamefont {W.~R.}\ \bibnamefont {Pla\ss{}}}, \bibinfo
  {author} {\bibfnamefont {D.}~\bibnamefont {Sch\"afer}}, \bibinfo {author}
  {\bibfnamefont {C.}~\bibnamefont {Scheidenberger}}, \bibinfo {author}
  {\bibfnamefont {J.~V.}\ \bibnamefont {de~Walle}}, \bibinfo {author}
  {\bibfnamefont {H.}~\bibnamefont {Weick}}, \ and\ \bibinfo {author}
  {\bibfnamefont {P.}~\bibnamefont {Dendooven}},\ }\href@noop {} {\bibfield
  {journal} {\bibinfo  {journal} {Europhys. Lett.}\ }\textbf {\bibinfo {volume}
  {96}},\ \bibinfo {pages} {52001} (\bibinfo {year} {2011})}\BibitemShut
  {NoStop}%
\bibitem [{\citenamefont {Purushothaman}\ \emph {et~al.}(2013)\citenamefont
  {Purushothaman}, \citenamefont {Reiter}, \citenamefont {Haettner},
  \citenamefont {Dendooven}, \citenamefont {Dickel}, \citenamefont {Geissel},
  \citenamefont {Ebert}, \citenamefont {Jesch}, \citenamefont {Pla\ss{}},
  \citenamefont {Ranjan}, \citenamefont {Weick}, \citenamefont {Amjad},
  \citenamefont {Ayet San~Andr\'es}, \citenamefont {Diwisch}, \citenamefont
  {Estrade}, \citenamefont {Farinon}, \citenamefont {Greiner}, \citenamefont
  {Kalantar-Nayestanaki}, \citenamefont {Kn\"obel}, \citenamefont {Kurcewicz},
  \citenamefont {Lang}, \citenamefont {Moore}, \citenamefont {Mukha},
  \citenamefont {Nociforo}, \citenamefont {Petrick}, \citenamefont
  {Pf\"utzner}, \citenamefont {Pietri}, \citenamefont {Prochazka},
  \citenamefont {Rink}, \citenamefont {Rinta-Antila}, \citenamefont
  {Scheidenberger}, \citenamefont {Takechi}, \citenamefont {Tanaka},
  \citenamefont {Winfield},\ and\ \citenamefont {Yavor}}]{Purushothaman2013}%
  \BibitemOpen
  \bibfield  {author} {\bibinfo {author} {\bibfnamefont {S.}~\bibnamefont
  {Purushothaman}}, \bibinfo {author} {\bibfnamefont {M.~P.}\ \bibnamefont
  {Reiter}}, \bibinfo {author} {\bibfnamefont {E.}~\bibnamefont {Haettner}},
  \bibinfo {author} {\bibfnamefont {P.}~\bibnamefont {Dendooven}}, \bibinfo
  {author} {\bibfnamefont {T.}~\bibnamefont {Dickel}}, \bibinfo {author}
  {\bibfnamefont {H.}~\bibnamefont {Geissel}}, \bibinfo {author} {\bibfnamefont
  {J.}~\bibnamefont {Ebert}}, \bibinfo {author} {\bibfnamefont
  {C.}~\bibnamefont {Jesch}}, \bibinfo {author} {\bibfnamefont {W.~R.}\
  \bibnamefont {Pla\ss{}}}, \bibinfo {author} {\bibfnamefont {M.}~\bibnamefont
  {Ranjan}}, \bibinfo {author} {\bibfnamefont {H.}~\bibnamefont {Weick}},
  \bibinfo {author} {\bibfnamefont {F.}~\bibnamefont {Amjad}}, \bibinfo
  {author} {\bibfnamefont {S.}~\bibnamefont {Ayet San~Andr\'es}}, \bibinfo
  {author} {\bibfnamefont {M.}~\bibnamefont {Diwisch}}, \bibinfo {author}
  {\bibfnamefont {A.}~\bibnamefont {Estrade}}, \bibinfo {author} {\bibfnamefont
  {F.}~\bibnamefont {Farinon}}, \bibinfo {author} {\bibfnamefont
  {F.}~\bibnamefont {Greiner}}, \bibinfo {author} {\bibfnamefont
  {N.}~\bibnamefont {Kalantar-Nayestanaki}}, \bibinfo {author} {\bibfnamefont
  {R.}~\bibnamefont {Kn\"obel}}, \bibinfo {author} {\bibfnamefont
  {J.}~\bibnamefont {Kurcewicz}}, \bibinfo {author} {\bibfnamefont
  {J.}~\bibnamefont {Lang}}, \bibinfo {author} {\bibfnamefont {I.~D.}\
  \bibnamefont {Moore}}, \bibinfo {author} {\bibfnamefont {I.}~\bibnamefont
  {Mukha}}, \bibinfo {author} {\bibfnamefont {C.}~\bibnamefont {Nociforo}},
  \bibinfo {author} {\bibfnamefont {M.}~\bibnamefont {Petrick}}, \bibinfo
  {author} {\bibfnamefont {M.}~\bibnamefont {Pf\"utzner}}, \bibinfo {author}
  {\bibfnamefont {S.}~\bibnamefont {Pietri}}, \bibinfo {author} {\bibfnamefont
  {A.}~\bibnamefont {Prochazka}}, \bibinfo {author} {\bibfnamefont {A.-K.}\
  \bibnamefont {Rink}}, \bibinfo {author} {\bibfnamefont {S.}~\bibnamefont
  {Rinta-Antila}}, \bibinfo {author} {\bibfnamefont {C.}~\bibnamefont
  {Scheidenberger}}, \bibinfo {author} {\bibfnamefont {M.}~\bibnamefont
  {Takechi}}, \bibinfo {author} {\bibfnamefont {Y.~K.}\ \bibnamefont {Tanaka}},
  \bibinfo {author} {\bibfnamefont {J.~S.}\ \bibnamefont {Winfield}}, \ and\
  \bibinfo {author} {\bibfnamefont {M.~I.}\ \bibnamefont {Yavor}},\ }\href@noop
  {} {\bibfield  {journal} {\bibinfo  {journal} {Europhys. Lett.}\ }\textbf
  {\bibinfo {volume} {104}},\ \bibinfo {pages} {42001} (\bibinfo {year}
  {2013})}\BibitemShut {NoStop}%
\bibitem [{\citenamefont {Ranjan}\ \emph {et~al.}(2015)\citenamefont {Ranjan},
  \citenamefont {Dendooven}, \citenamefont {Purushothaman}, \citenamefont
  {Dickel}, \citenamefont {Reiter}, \citenamefont {Ayet San~Andr\'es},
  \citenamefont {Haettner}, \citenamefont {Moore}, \citenamefont
  {Kalantar-Nayestanaki}, \citenamefont {Geissel}, \citenamefont {Pla\ss{}},
  \citenamefont {Sch\"afer}, \citenamefont {Scheidenberger}, \citenamefont
  {Schreuder}, \citenamefont {Timersma}, \citenamefont {de~Walle},\ and\
  \citenamefont {Weick}}]{Ranjan2015}%
  \BibitemOpen
  \bibfield  {author} {\bibinfo {author} {\bibfnamefont {M.}~\bibnamefont
  {Ranjan}}, \bibinfo {author} {\bibfnamefont {P.}~\bibnamefont {Dendooven}},
  \bibinfo {author} {\bibfnamefont {S.}~\bibnamefont {Purushothaman}}, \bibinfo
  {author} {\bibfnamefont {T.}~\bibnamefont {Dickel}}, \bibinfo {author}
  {\bibfnamefont {M.}~\bibnamefont {Reiter}}, \bibinfo {author} {\bibfnamefont
  {S.}~\bibnamefont {Ayet San~Andr\'es}}, \bibinfo {author} {\bibfnamefont
  {E.}~\bibnamefont {Haettner}}, \bibinfo {author} {\bibfnamefont
  {I.}~\bibnamefont {Moore}}, \bibinfo {author} {\bibfnamefont
  {N.}~\bibnamefont {Kalantar-Nayestanaki}}, \bibinfo {author} {\bibfnamefont
  {H.}~\bibnamefont {Geissel}}, \bibinfo {author} {\bibfnamefont
  {W.}~\bibnamefont {Pla\ss{}}}, \bibinfo {author} {\bibfnamefont
  {D.}~\bibnamefont {Sch\"afer}}, \bibinfo {author} {\bibfnamefont
  {C.}~\bibnamefont {Scheidenberger}}, \bibinfo {author} {\bibfnamefont
  {F.}~\bibnamefont {Schreuder}}, \bibinfo {author} {\bibfnamefont
  {H.}~\bibnamefont {Timersma}}, \bibinfo {author} {\bibfnamefont {J.~V.}\
  \bibnamefont {de~Walle}}, \ and\ \bibinfo {author} {\bibfnamefont
  {H.}~\bibnamefont {Weick}},\ }\href@noop {} {\bibfield  {journal} {\bibinfo
  {journal} {Nucl. Instrum. Methods A}\ }\textbf {\bibinfo {volume} {770}},\
  \bibinfo {pages} {87 } (\bibinfo {year} {2015})}\BibitemShut {NoStop}%
\bibitem [{\citenamefont {Greiner}\ \emph {et~al.}(2019)\citenamefont
  {Greiner}, \citenamefont {Dickel}, \citenamefont {Ayet}, \citenamefont
  {Bergmann}, \citenamefont {Constantin}, \citenamefont {Ebert}, \citenamefont
  {Geissel}, \citenamefont {Haettner}, \citenamefont {Hornung}, \citenamefont
  {Miskun}, \citenamefont {Lippert}, \citenamefont {Mardor}, \citenamefont
  {Moore}, \citenamefont {Pla\ss{}}, \citenamefont {Purushothaman},
  \citenamefont {Rink}, \citenamefont {Reiter}, \citenamefont
  {Scheidenberger},\ and\ \citenamefont {Weick}}]{Greiner2019}%
  \BibitemOpen
  \bibfield  {author} {\bibinfo {author} {\bibfnamefont {F.}~\bibnamefont
  {Greiner}}, \bibinfo {author} {\bibfnamefont {T.}~\bibnamefont {Dickel}},
  \bibinfo {author} {\bibfnamefont {S.}~\bibnamefont {Ayet}}, \bibinfo {author}
  {\bibfnamefont {J.}~\bibnamefont {Bergmann}}, \bibinfo {author}
  {\bibfnamefont {P.}~\bibnamefont {Constantin}}, \bibinfo {author}
  {\bibfnamefont {J.}~\bibnamefont {Ebert}}, \bibinfo {author} {\bibfnamefont
  {H.}~\bibnamefont {Geissel}}, \bibinfo {author} {\bibfnamefont
  {E.}~\bibnamefont {Haettner}}, \bibinfo {author} {\bibfnamefont
  {C.}~\bibnamefont {Hornung}}, \bibinfo {author} {\bibfnamefont
  {I.}~\bibnamefont {Miskun}}, \bibinfo {author} {\bibfnamefont
  {W.}~\bibnamefont {Lippert}}, \bibinfo {author} {\bibfnamefont
  {I.}~\bibnamefont {Mardor}}, \bibinfo {author} {\bibfnamefont
  {I.}~\bibnamefont {Moore}}, \bibinfo {author} {\bibfnamefont {W.~R.}\
  \bibnamefont {Pla\ss{}}}, \bibinfo {author} {\bibfnamefont {S.}~\bibnamefont
  {Purushothaman}}, \bibinfo {author} {\bibfnamefont {A.-K.}\ \bibnamefont
  {Rink}}, \bibinfo {author} {\bibfnamefont {M.~P.}\ \bibnamefont {Reiter}},
  \bibinfo {author} {\bibfnamefont {C.}~\bibnamefont {Scheidenberger}}, \ and\
  \bibinfo {author} {\bibfnamefont {H.}~\bibnamefont {Weick}},\ }\href
  {\doibase https://doi.org/10.1016/j.nimb.2019.04.072} {\bibfield  {journal}
  {\bibinfo  {journal} {Nucl. Instrum. Methods Phys. Res. B}\ } (\bibinfo
  {year} {2019}),\ https://doi.org/10.1016/j.nimb.2019.04.072}\BibitemShut
  {NoStop}%
\bibitem [{\citenamefont {Haettner}\ \emph {et~al.}(2018)\citenamefont
  {Haettner}, \citenamefont {Pla\ss{}}, \citenamefont {Czok}, \citenamefont
  {Dickel}, \citenamefont {Geissel}, \citenamefont {Kinsel}, \citenamefont
  {Petrick}, \citenamefont {Sch\"afer},\ and\ \citenamefont
  {Scheidenberger}}]{Haettner2018}%
  \BibitemOpen
  \bibfield  {author} {\bibinfo {author} {\bibfnamefont {E.}~\bibnamefont
  {Haettner}}, \bibinfo {author} {\bibfnamefont {W.~R.}\ \bibnamefont
  {Pla\ss{}}}, \bibinfo {author} {\bibfnamefont {U.}~\bibnamefont {Czok}},
  \bibinfo {author} {\bibfnamefont {T.}~\bibnamefont {Dickel}}, \bibinfo
  {author} {\bibfnamefont {H.}~\bibnamefont {Geissel}}, \bibinfo {author}
  {\bibfnamefont {W.}~\bibnamefont {Kinsel}}, \bibinfo {author} {\bibfnamefont
  {M.}~\bibnamefont {Petrick}}, \bibinfo {author} {\bibfnamefont
  {T.}~\bibnamefont {Sch\"afer}}, \ and\ \bibinfo {author} {\bibfnamefont
  {C.}~\bibnamefont {Scheidenberger}},\ }\href@noop {} {\bibfield  {journal}
  {\bibinfo  {journal} {Nucl. Inst. Methods A}\ }\textbf {\bibinfo {volume}
  {880}} (\bibinfo {year} {2018})}\BibitemShut {NoStop}%
\bibitem [{\citenamefont {Pla\ss{}}\ \emph {et~al.}(2008)\citenamefont
  {Pla\ss{}}, \citenamefont {Dickel}, \citenamefont {Czok}, \citenamefont
  {Geissel}, \citenamefont {Petrick}, \citenamefont {Reinheimer}, \citenamefont
  {Scheidenberger},\ and\ \citenamefont {Yavor}}]{Plass2008}%
  \BibitemOpen
  \bibfield  {author} {\bibinfo {author} {\bibfnamefont {W.~R.}\ \bibnamefont
  {Pla\ss{}}}, \bibinfo {author} {\bibfnamefont {T.}~\bibnamefont {Dickel}},
  \bibinfo {author} {\bibfnamefont {U.}~\bibnamefont {Czok}}, \bibinfo {author}
  {\bibfnamefont {H.}~\bibnamefont {Geissel}}, \bibinfo {author} {\bibfnamefont
  {M.}~\bibnamefont {Petrick}}, \bibinfo {author} {\bibfnamefont
  {K.}~\bibnamefont {Reinheimer}}, \bibinfo {author} {\bibfnamefont
  {C.}~\bibnamefont {Scheidenberger}}, \ and\ \bibinfo {author} {\bibfnamefont
  {M.}~\bibnamefont {Yavor}},\ }\href@noop {} {\bibfield  {journal} {\bibinfo
  {journal} {Nucl. Instrum. Methods B}\ }\textbf {\bibinfo {volume} {266}},\
  \bibinfo {pages} {4560} (\bibinfo {year} {2008})}\BibitemShut {NoStop}%
\bibitem [{\citenamefont {Dickel}\ \emph
  {et~al.}(2015{\natexlab{b}})\citenamefont {Dickel}, \citenamefont {Pla\ss{}},
  \citenamefont {Becker}, \citenamefont {Czok}, \citenamefont {Geissel},
  \citenamefont {Haettner}, \citenamefont {Jesch}, \citenamefont {Kinsel},
  \citenamefont {Petrick}, \citenamefont {Scheidenberger}, \citenamefont
  {Simon},\ and\ \citenamefont {Yavor}}]{Dickel2015b}%
  \BibitemOpen
  \bibfield  {author} {\bibinfo {author} {\bibfnamefont {T.}~\bibnamefont
  {Dickel}}, \bibinfo {author} {\bibfnamefont {W.~R.}\ \bibnamefont
  {Pla\ss{}}}, \bibinfo {author} {\bibfnamefont {A.}~\bibnamefont {Becker}},
  \bibinfo {author} {\bibfnamefont {U.}~\bibnamefont {Czok}}, \bibinfo {author}
  {\bibfnamefont {H.}~\bibnamefont {Geissel}}, \bibinfo {author} {\bibfnamefont
  {E.}~\bibnamefont {Haettner}}, \bibinfo {author} {\bibfnamefont
  {C.}~\bibnamefont {Jesch}}, \bibinfo {author} {\bibfnamefont
  {W.}~\bibnamefont {Kinsel}}, \bibinfo {author} {\bibfnamefont
  {M.}~\bibnamefont {Petrick}}, \bibinfo {author} {\bibfnamefont
  {C.}~\bibnamefont {Scheidenberger}}, \bibinfo {author} {\bibfnamefont
  {A.}~\bibnamefont {Simon}}, \ and\ \bibinfo {author} {\bibfnamefont {M.~I.}\
  \bibnamefont {Yavor}},\ }\href@noop {} {\bibfield  {journal} {\bibinfo
  {journal} {Nucl. Instrum. Methods A}\ }\textbf {\bibinfo {volume} {{777}}},\
  \bibinfo {pages} {172} (\bibinfo {year} {{2015}}{\natexlab{b}})}\BibitemShut
  {NoStop}%
\bibitem [{\citenamefont {Will}(2019)}]{will2019}%
  \BibitemOpen
  \bibfield  {author} {\bibinfo {author} {\bibfnamefont {C.}~\bibnamefont
  {Will}},\ }\emph {\bibinfo {title} {Achieving one million mass resolving
  power with a multiple-reflection time-of-flight mass spectrometer}},\
  \href@noop {} {\bibinfo {type} {M.{S}c. thesis}},\ \bibinfo  {school} {Justus
  Liebig Universit\"at Gie\ss{}en} (\bibinfo {year} {2019})\BibitemShut
  {NoStop}%
\bibitem [{\citenamefont {Beck}(2021)}]{Beck2020}%
  \BibitemOpen
  \bibfield  {author} {\bibinfo {author} {\bibfnamefont {S.}~\bibnamefont
  {Beck}},\ }\emph {\bibinfo {title} {High resolving power mass measurements
  with a multiple-reflection time-of-flight mass spectrometer - in
  preparation}},\ \href@noop {} {\bibinfo {type} {Ph.{D}. thesis}},\ \bibinfo
  {school} {Justus Liebig Universit\"at Gie\ss{}en} (\bibinfo {year}
  {2021})\BibitemShut {NoStop}%
\bibitem [{\citenamefont {Gr\"of}(2020)}]{Groef2020}%
  \BibitemOpen
  \bibfield  {author} {\bibinfo {author} {\bibfnamefont {L.}~\bibnamefont
  {Gr\"of}},\ }\emph {\bibinfo {title} {Range bunching at the FRS}},\
  \href@noop {} {\bibinfo {type} {M.{S}c. thesis}},\ \bibinfo  {school} {Justus
  Liebig Universit\"at Gie\ss{}en} (\bibinfo {year} {2020})\BibitemShut
  {NoStop}%
\bibitem [{\citenamefont {Mukha}\ \emph {et~al.}(2018)\citenamefont {Mukha},
  \citenamefont {Grigorenko}, \citenamefont {Kostyleva}, \citenamefont
  {Acosta}, \citenamefont {Casarejos}, \citenamefont {Ciemny}, \citenamefont
  {Dominik}, \citenamefont {Due\~nas}, \citenamefont {Dunin}, \citenamefont
  {Espino}, \citenamefont {Estrad\'e}, \citenamefont {Farinon}, \citenamefont
  {Fomichev}, \citenamefont {Geissel}, \citenamefont {Gorshkov}, \citenamefont
  {Janas}, \citenamefont {Kami\ifmmode~\acute{n}\else \'{n}\fi{}ski},
  \citenamefont {Kiselev}, \citenamefont {Kn\"obel}, \citenamefont {Krupko},
  \citenamefont {Kuich}, \citenamefont {Litvinov}, \citenamefont
  {Marquinez-Dur\'an}, \citenamefont {Martel}, \citenamefont {Mazzocchi},
  \citenamefont {Nociforo}, \citenamefont {Ord\'uz}, \citenamefont
  {Pf\"utzner}, \citenamefont {Pietri}, \citenamefont {Pomorski}, \citenamefont
  {Prochazka}, \citenamefont {Rymzhanova}, \citenamefont
  {S\'anchez-Ben\'{\i}tez}, \citenamefont {Scheidenberger}, \citenamefont
  {Sharov}, \citenamefont {Simon}, \citenamefont {Sitar}, \citenamefont
  {Slepnev}, \citenamefont {Stanoiu}, \citenamefont {Strmen}, \citenamefont
  {Szarka}, \citenamefont {Takechi}, \citenamefont {Tanaka}, \citenamefont
  {Weick}, \citenamefont {Winkler}, \citenamefont {Winfield}, \citenamefont
  {Xu},\ and\ \citenamefont {Zhukov}}]{Mukha2018}%
  \BibitemOpen
  \bibfield  {author} {\bibinfo {author} {\bibfnamefont {I.}~\bibnamefont
  {Mukha}}, \bibinfo {author} {\bibfnamefont {L.~V.}\ \bibnamefont
  {Grigorenko}}, \bibinfo {author} {\bibfnamefont {D.}~\bibnamefont
  {Kostyleva}}, \bibinfo {author} {\bibfnamefont {L.}~\bibnamefont {Acosta}},
  \bibinfo {author} {\bibfnamefont {E.}~\bibnamefont {Casarejos}}, \bibinfo
  {author} {\bibfnamefont {A.~A.}\ \bibnamefont {Ciemny}}, \bibinfo {author}
  {\bibfnamefont {W.}~\bibnamefont {Dominik}}, \bibinfo {author} {\bibfnamefont
  {J.~A.}\ \bibnamefont {Due\~nas}}, \bibinfo {author} {\bibfnamefont
  {V.}~\bibnamefont {Dunin}}, \bibinfo {author} {\bibfnamefont {J.~M.}\
  \bibnamefont {Espino}}, \bibinfo {author} {\bibfnamefont {A.}~\bibnamefont
  {Estrad\'e}}, \bibinfo {author} {\bibfnamefont {F.}~\bibnamefont {Farinon}},
  \bibinfo {author} {\bibfnamefont {A.}~\bibnamefont {Fomichev}}, \bibinfo
  {author} {\bibfnamefont {H.}~\bibnamefont {Geissel}}, \bibinfo {author}
  {\bibfnamefont {A.}~\bibnamefont {Gorshkov}}, \bibinfo {author}
  {\bibfnamefont {Z.}~\bibnamefont {Janas}}, \bibinfo {author} {\bibfnamefont
  {G.}~\bibnamefont {Kami\ifmmode~\acute{n}\else \'{n}\fi{}ski}}, \bibinfo
  {author} {\bibfnamefont {O.}~\bibnamefont {Kiselev}}, \bibinfo {author}
  {\bibfnamefont {R.}~\bibnamefont {Kn\"obel}}, \bibinfo {author}
  {\bibfnamefont {S.}~\bibnamefont {Krupko}}, \bibinfo {author} {\bibfnamefont
  {M.}~\bibnamefont {Kuich}}, \bibinfo {author} {\bibfnamefont {Y.~A.}\
  \bibnamefont {Litvinov}}, \bibinfo {author} {\bibfnamefont {G.}~\bibnamefont
  {Marquinez-Dur\'an}}, \bibinfo {author} {\bibfnamefont {I.}~\bibnamefont
  {Martel}}, \bibinfo {author} {\bibfnamefont {C.}~\bibnamefont {Mazzocchi}},
  \bibinfo {author} {\bibfnamefont {C.}~\bibnamefont {Nociforo}}, \bibinfo
  {author} {\bibfnamefont {A.~K.}\ \bibnamefont {Ord\'uz}}, \bibinfo {author}
  {\bibfnamefont {M.}~\bibnamefont {Pf\"utzner}}, \bibinfo {author}
  {\bibfnamefont {S.}~\bibnamefont {Pietri}}, \bibinfo {author} {\bibfnamefont
  {M.}~\bibnamefont {Pomorski}}, \bibinfo {author} {\bibfnamefont
  {A.}~\bibnamefont {Prochazka}}, \bibinfo {author} {\bibfnamefont
  {S.}~\bibnamefont {Rymzhanova}}, \bibinfo {author} {\bibfnamefont {A.~M.}\
  \bibnamefont {S\'anchez-Ben\'{\i}tez}}, \bibinfo {author} {\bibfnamefont
  {C.}~\bibnamefont {Scheidenberger}}, \bibinfo {author} {\bibfnamefont
  {P.}~\bibnamefont {Sharov}}, \bibinfo {author} {\bibfnamefont
  {H.}~\bibnamefont {Simon}}, \bibinfo {author} {\bibfnamefont
  {B.}~\bibnamefont {Sitar}}, \bibinfo {author} {\bibfnamefont
  {R.}~\bibnamefont {Slepnev}}, \bibinfo {author} {\bibfnamefont
  {M.}~\bibnamefont {Stanoiu}}, \bibinfo {author} {\bibfnamefont
  {P.}~\bibnamefont {Strmen}}, \bibinfo {author} {\bibfnamefont
  {I.}~\bibnamefont {Szarka}}, \bibinfo {author} {\bibfnamefont
  {M.}~\bibnamefont {Takechi}}, \bibinfo {author} {\bibfnamefont {Y.~K.}\
  \bibnamefont {Tanaka}}, \bibinfo {author} {\bibfnamefont {H.}~\bibnamefont
  {Weick}}, \bibinfo {author} {\bibfnamefont {M.}~\bibnamefont {Winkler}},
  \bibinfo {author} {\bibfnamefont {J.~S.}\ \bibnamefont {Winfield}}, \bibinfo
  {author} {\bibfnamefont {X.}~\bibnamefont {Xu}}, \ and\ \bibinfo {author}
  {\bibfnamefont {M.~V.}\ \bibnamefont {Zhukov}},\ }\href {\doibase
  10.1103/PhysRevC.98.064308} {\bibfield  {journal} {\bibinfo  {journal} {Phys.
  Rev. C}\ }\textbf {\bibinfo {volume} {98}},\ \bibinfo {pages} {064308}
  (\bibinfo {year} {2018})}\BibitemShut {NoStop}%
\bibitem [{\citenamefont {Purushothaman}\ \emph {et~al.}(2017)\citenamefont
  {Purushothaman}, \citenamefont {Ayet San~Andr\'es}, \citenamefont {Bergmann},
  \citenamefont {Dickel}, \citenamefont {Ebert}, \citenamefont {Geissel},
  \citenamefont {Hornung}, \citenamefont {Pla\ss{}}, \citenamefont {Rappold},
  \citenamefont {Scheidenberger}, \citenamefont {Tanaka},\ and\ \citenamefont
  {Yavor}}]{Purushothaman2017}%
  \BibitemOpen
  \bibfield  {author} {\bibinfo {author} {\bibfnamefont {S.}~\bibnamefont
  {Purushothaman}}, \bibinfo {author} {\bibfnamefont {S.}~\bibnamefont {Ayet
  San~Andr\'es}}, \bibinfo {author} {\bibfnamefont {J.}~\bibnamefont
  {Bergmann}}, \bibinfo {author} {\bibfnamefont {T.}~\bibnamefont {Dickel}},
  \bibinfo {author} {\bibfnamefont {J.}~\bibnamefont {Ebert}}, \bibinfo
  {author} {\bibfnamefont {H.}~\bibnamefont {Geissel}}, \bibinfo {author}
  {\bibfnamefont {C.}~\bibnamefont {Hornung}}, \bibinfo {author} {\bibfnamefont
  {W.~R.}\ \bibnamefont {Pla\ss{}}}, \bibinfo {author} {\bibfnamefont
  {C.}~\bibnamefont {Rappold}}, \bibinfo {author} {\bibfnamefont
  {C.}~\bibnamefont {Scheidenberger}}, \bibinfo {author} {\bibfnamefont
  {Y.~K.}\ \bibnamefont {Tanaka}}, \ and\ \bibinfo {author} {\bibfnamefont
  {M.~I.}\ \bibnamefont {Yavor}},\ }\href@noop {} {\bibfield  {journal}
  {\bibinfo  {journal} {Int. J. Mass Spectrom.}\ }\textbf {\bibinfo {volume}
  {{421}}},\ \bibinfo {pages} {245} (\bibinfo {year} {{2017}})}\BibitemShut
  {NoStop}%
\bibitem [{\citenamefont {Boswell}\ and\ \citenamefont
  {McGee}(1970)}]{Boswell1970}%
  \BibitemOpen
  \bibfield  {author} {\bibinfo {author} {\bibfnamefont {G.}~\bibnamefont
  {Boswell}}\ and\ \bibinfo {author} {\bibfnamefont {T.}~\bibnamefont
  {McGee}},\ }\href {\doibase https://doi.org/10.1016/0022-1902(70)80338-4}
  {\bibfield  {journal} {\bibinfo  {journal} {J. Inorg. Nucl. Chem.}\ }\textbf
  {\bibinfo {volume} {32}},\ \bibinfo {pages} {2805 } (\bibinfo {year}
  {1970})}\BibitemShut {NoStop}%
\bibitem [{\citenamefont {Macdonald}\ \emph {et~al.}(1977)\citenamefont
  {Macdonald}, \citenamefont {Hardy}, \citenamefont {Schmeing}, \citenamefont
  {Faestermann}, \citenamefont {Andrews}, \citenamefont {Geiger}, \citenamefont
  {Graham},\ and\ \citenamefont {Jackson}}]{Macdonald1977}%
  \BibitemOpen
  \bibfield  {author} {\bibinfo {author} {\bibfnamefont {J.}~\bibnamefont
  {Macdonald}}, \bibinfo {author} {\bibfnamefont {J.}~\bibnamefont {Hardy}},
  \bibinfo {author} {\bibfnamefont {H.}~\bibnamefont {Schmeing}}, \bibinfo
  {author} {\bibfnamefont {T.}~\bibnamefont {Faestermann}}, \bibinfo {author}
  {\bibfnamefont {H.}~\bibnamefont {Andrews}}, \bibinfo {author} {\bibfnamefont
  {J.}~\bibnamefont {Geiger}}, \bibinfo {author} {\bibfnamefont
  {R.}~\bibnamefont {Graham}}, \ and\ \bibinfo {author} {\bibfnamefont
  {K.}~\bibnamefont {Jackson}},\ }\href {\doibase
  https://doi.org/10.1016/0375-9474(77)90078-1} {\bibfield  {journal} {\bibinfo
   {journal} {Nucl. Phys. A}\ }\textbf {\bibinfo {volume} {288}},\ \bibinfo
  {pages} {1 } (\bibinfo {year} {1977})}\BibitemShut {NoStop}%
\bibitem [{\citenamefont {Huang}\ \emph {et~al.}(2017)\citenamefont {Huang},
  \citenamefont {Audi}, \citenamefont {Wang}, \citenamefont {Kondev},
  \citenamefont {Naimi},\ and\ \citenamefont {Xu}}]{AME16}%
  \BibitemOpen
  \bibfield  {author} {\bibinfo {author} {\bibfnamefont {W.}~\bibnamefont
  {Huang}}, \bibinfo {author} {\bibfnamefont {G.}~\bibnamefont {Audi}},
  \bibinfo {author} {\bibfnamefont {M.}~\bibnamefont {Wang}}, \bibinfo {author}
  {\bibfnamefont {F.}~\bibnamefont {Kondev}}, \bibinfo {author} {\bibfnamefont
  {S.}~\bibnamefont {Naimi}}, \ and\ \bibinfo {author} {\bibfnamefont
  {X.}~\bibnamefont {Xu}},\ }\href
  {http://stacks.iop.org/1674-1137/41/i=3/a=030002} {\bibfield  {journal}
  {\bibinfo  {journal} {Chinese Phys. C}\ }\textbf {\bibinfo {volume} {41}},\
  \bibinfo {pages} {030002} (\bibinfo {year} {2017})}\BibitemShut {NoStop}%
\bibitem [{\citenamefont {Lima}\ \emph {et~al.}(2002)\citenamefont {Lima},
  \citenamefont {L\'epine-Szily}, \citenamefont {Audi}, \citenamefont {Mittig},
  \citenamefont {Chartier}, \citenamefont {Orr}, \citenamefont {Lichtenthaler},
  \citenamefont {Angelique}, \citenamefont {Casandjian}, \citenamefont
  {Cunsolo}, \citenamefont {Donzaud}, \citenamefont {Foti}, \citenamefont
  {Gillibert}, \citenamefont {Lewitowicz}, \citenamefont {Lukyanov},
  \citenamefont {MacCormick}, \citenamefont {Morrissey}, \citenamefont
  {Ostrowski}, \citenamefont {Sherrill}, \citenamefont {Stephan}, \citenamefont
  {Suomijarvi}, \citenamefont {Tassan-Got}, \citenamefont {Vieira},
  \citenamefont {Villari},\ and\ \citenamefont {Wouters}}]{Lima2002}%
  \BibitemOpen
  \bibfield  {author} {\bibinfo {author} {\bibfnamefont {G.~F.}\ \bibnamefont
  {Lima}}, \bibinfo {author} {\bibfnamefont {A.}~\bibnamefont
  {L\'epine-Szily}}, \bibinfo {author} {\bibfnamefont {G.}~\bibnamefont
  {Audi}}, \bibinfo {author} {\bibfnamefont {W.}~\bibnamefont {Mittig}},
  \bibinfo {author} {\bibfnamefont {M.}~\bibnamefont {Chartier}}, \bibinfo
  {author} {\bibfnamefont {N.~A.}\ \bibnamefont {Orr}}, \bibinfo {author}
  {\bibfnamefont {R.}~\bibnamefont {Lichtenthaler}}, \bibinfo {author}
  {\bibfnamefont {J.~C.}\ \bibnamefont {Angelique}}, \bibinfo {author}
  {\bibfnamefont {J.~M.}\ \bibnamefont {Casandjian}}, \bibinfo {author}
  {\bibfnamefont {A.}~\bibnamefont {Cunsolo}}, \bibinfo {author} {\bibfnamefont
  {C.}~\bibnamefont {Donzaud}}, \bibinfo {author} {\bibfnamefont
  {A.}~\bibnamefont {Foti}}, \bibinfo {author} {\bibfnamefont {A.}~\bibnamefont
  {Gillibert}}, \bibinfo {author} {\bibfnamefont {M.}~\bibnamefont
  {Lewitowicz}}, \bibinfo {author} {\bibfnamefont {S.}~\bibnamefont
  {Lukyanov}}, \bibinfo {author} {\bibfnamefont {M.}~\bibnamefont
  {MacCormick}}, \bibinfo {author} {\bibfnamefont {D.~J.}\ \bibnamefont
  {Morrissey}}, \bibinfo {author} {\bibfnamefont {A.~N.}\ \bibnamefont
  {Ostrowski}}, \bibinfo {author} {\bibfnamefont {B.~M.}\ \bibnamefont
  {Sherrill}}, \bibinfo {author} {\bibfnamefont {C.}~\bibnamefont {Stephan}},
  \bibinfo {author} {\bibfnamefont {T.}~\bibnamefont {Suomijarvi}}, \bibinfo
  {author} {\bibfnamefont {L.}~\bibnamefont {Tassan-Got}}, \bibinfo {author}
  {\bibfnamefont {D.~J.}\ \bibnamefont {Vieira}}, \bibinfo {author}
  {\bibfnamefont {A.~C.~C.}\ \bibnamefont {Villari}}, \ and\ \bibinfo {author}
  {\bibfnamefont {J.~M.}\ \bibnamefont {Wouters}},\ }\href {\doibase
  10.1103/PhysRevC.65.044618} {\bibfield  {journal} {\bibinfo  {journal} {Phys.
  Rev. C}\ }\textbf {\bibinfo {volume} {65}},\ \bibinfo {pages} {044618}
  (\bibinfo {year} {2002})}\BibitemShut {NoStop}%
\bibitem [{\citenamefont {Chartier}\ \emph {et~al.}(1998)\citenamefont
  {Chartier}, \citenamefont {Mittig}, \citenamefont {Orr}, \citenamefont
  {Angélique}, \citenamefont {Audi}, \citenamefont {Casandjian}, \citenamefont
  {Cunsolo}, \citenamefont {Donzaud}, \citenamefont {Foti}, \citenamefont
  {Lépine-Szily}, \citenamefont {Lewitowicz}, \citenamefont {Lukyanov},
  \citenamefont {{Mac Cormick}}, \citenamefont {Morrissey}, \citenamefont
  {Ostrowski}, \citenamefont {Sherrill}, \citenamefont {Stephan}, \citenamefont
  {Suomijärvi}, \citenamefont {Tassan-Got}, \citenamefont {Vieira},
  \citenamefont {Villari},\ and\ \citenamefont {Wouters}}]{Chartier1998}%
  \BibitemOpen
  \bibfield  {author} {\bibinfo {author} {\bibfnamefont {M.}~\bibnamefont
  {Chartier}}, \bibinfo {author} {\bibfnamefont {W.}~\bibnamefont {Mittig}},
  \bibinfo {author} {\bibfnamefont {N.}~\bibnamefont {Orr}}, \bibinfo {author}
  {\bibfnamefont {J.-C.}\ \bibnamefont {Angélique}}, \bibinfo {author}
  {\bibfnamefont {G.}~\bibnamefont {Audi}}, \bibinfo {author} {\bibfnamefont
  {J.-M.}\ \bibnamefont {Casandjian}}, \bibinfo {author} {\bibfnamefont
  {A.}~\bibnamefont {Cunsolo}}, \bibinfo {author} {\bibfnamefont
  {C.}~\bibnamefont {Donzaud}}, \bibinfo {author} {\bibfnamefont
  {A.}~\bibnamefont {Foti}}, \bibinfo {author} {\bibfnamefont {A.}~\bibnamefont
  {Lépine-Szily}}, \bibinfo {author} {\bibfnamefont {M.}~\bibnamefont
  {Lewitowicz}}, \bibinfo {author} {\bibfnamefont {S.}~\bibnamefont
  {Lukyanov}}, \bibinfo {author} {\bibfnamefont {M.}~\bibnamefont {{Mac
  Cormick}}}, \bibinfo {author} {\bibfnamefont {D.}~\bibnamefont {Morrissey}},
  \bibinfo {author} {\bibfnamefont {A.}~\bibnamefont {Ostrowski}}, \bibinfo
  {author} {\bibfnamefont {B.}~\bibnamefont {Sherrill}}, \bibinfo {author}
  {\bibfnamefont {C.}~\bibnamefont {Stephan}}, \bibinfo {author} {\bibfnamefont
  {T.}~\bibnamefont {Suomijärvi}}, \bibinfo {author} {\bibfnamefont
  {L.}~\bibnamefont {Tassan-Got}}, \bibinfo {author} {\bibfnamefont
  {D.}~\bibnamefont {Vieira}}, \bibinfo {author} {\bibfnamefont
  {A.}~\bibnamefont {Villari}}, \ and\ \bibinfo {author} {\bibfnamefont
  {J.}~\bibnamefont {Wouters}},\ }\href {\doibase
  https://doi.org/10.1016/S0375-9474(98)00228-0} {\bibfield  {journal}
  {\bibinfo  {journal} {Nucl. Phys. A}\ }\textbf {\bibinfo {volume} {637}},\
  \bibinfo {pages} {3 } (\bibinfo {year} {1998})}\BibitemShut {NoStop}%
\bibitem [{\citenamefont {Herfurth}\ \emph {et~al.}(2011)\citenamefont
  {Herfurth}, \citenamefont {Audi}, \citenamefont {Beck}, \citenamefont
  {Blaum}, \citenamefont {Bollen}, \citenamefont {Delahaye}, \citenamefont
  {Dworschak}, \citenamefont {George}, \citenamefont {Guénaut}, \citenamefont
  {Kellerbauer}, \citenamefont {Lunney}, \citenamefont {Mukherjee},
  \citenamefont {Rahaman}, \citenamefont {Schwarz}, \citenamefont
  {Schweikhard}, \citenamefont {Weber},\ and\ \citenamefont
  {Yazidjian}}]{Herfurth2011}%
  \BibitemOpen
  \bibfield  {author} {\bibinfo {author} {\bibfnamefont {F.}~\bibnamefont
  {Herfurth}}, \bibinfo {author} {\bibfnamefont {G.}~\bibnamefont {Audi}},
  \bibinfo {author} {\bibfnamefont {D.}~\bibnamefont {Beck}}, \bibinfo {author}
  {\bibfnamefont {K.}~\bibnamefont {Blaum}}, \bibinfo {author} {\bibfnamefont
  {G.}~\bibnamefont {Bollen}}, \bibinfo {author} {\bibfnamefont
  {P.}~\bibnamefont {Delahaye}}, \bibinfo {author} {\bibfnamefont
  {M.}~\bibnamefont {Dworschak}}, \bibinfo {author} {\bibfnamefont
  {S.}~\bibnamefont {George}}, \bibinfo {author} {\bibfnamefont
  {C.}~\bibnamefont {Guénaut}}, \bibinfo {author} {\bibfnamefont
  {A.}~\bibnamefont {Kellerbauer}}, \bibinfo {author} {\bibfnamefont
  {D.}~\bibnamefont {Lunney}}, \bibinfo {author} {\bibfnamefont
  {M.}~\bibnamefont {Mukherjee}}, \bibinfo {author} {\bibfnamefont
  {S.}~\bibnamefont {Rahaman}}, \bibinfo {author} {\bibfnamefont
  {S.}~\bibnamefont {Schwarz}}, \bibinfo {author} {\bibfnamefont
  {L.}~\bibnamefont {Schweikhard}}, \bibinfo {author} {\bibfnamefont
  {C.}~\bibnamefont {Weber}}, \ and\ \bibinfo {author} {\bibfnamefont
  {C.}~\bibnamefont {Yazidjian}},\ }\href {\doibase
  https://doi.org/10.1140/epja/i2011-11075-6} {\bibfield  {journal} {\bibinfo
  {journal} {Eur. Phys. Jou. A}\ }\textbf {\bibinfo {volume} {47}},\ \bibinfo
  {pages} {75} (\bibinfo {year} {2011})}\BibitemShut {NoStop}%
\bibitem [{\citenamefont {Hausmann}\ \emph {et~al.}(2001)\citenamefont
  {Hausmann}, \citenamefont {Stadlmann}, \citenamefont {Attallah},
  \citenamefont {Beckert}, \citenamefont {Beller}, \citenamefont {Bosch},
  \citenamefont {Eickhoff}, \citenamefont {Falch}, \citenamefont {Franczak},
  \citenamefont {Franzke}, \citenamefont {Geissel}, \citenamefont {Kerscher},
  \citenamefont {Klepper}, \citenamefont {Kluge}, \citenamefont {Kozhuharov},
  \citenamefont {Litvinov}, \citenamefont {Löbner}, \citenamefont
  {Münzenberg}, \citenamefont {Nankov}, \citenamefont {Nolden}, \citenamefont
  {Novikov}, \citenamefont {Ohtsubo}, \citenamefont {Radon}, \citenamefont
  {Schatz}, \citenamefont {Scheidenberger}, \citenamefont {Steck},
  \citenamefont {Sun}, \citenamefont {Weick},\ and\ \citenamefont
  {Wollnik}}]{Hausmann2001}%
  \BibitemOpen
  \bibfield  {author} {\bibinfo {author} {\bibfnamefont {M.}~\bibnamefont
  {Hausmann}}, \bibinfo {author} {\bibfnamefont {J.}~\bibnamefont {Stadlmann}},
  \bibinfo {author} {\bibfnamefont {F.}~\bibnamefont {Attallah}}, \bibinfo
  {author} {\bibfnamefont {K.}~\bibnamefont {Beckert}}, \bibinfo {author}
  {\bibfnamefont {P.}~\bibnamefont {Beller}}, \bibinfo {author} {\bibfnamefont
  {F.}~\bibnamefont {Bosch}}, \bibinfo {author} {\bibfnamefont
  {H.}~\bibnamefont {Eickhoff}}, \bibinfo {author} {\bibfnamefont
  {M.}~\bibnamefont {Falch}}, \bibinfo {author} {\bibfnamefont
  {B.}~\bibnamefont {Franczak}}, \bibinfo {author} {\bibfnamefont
  {B.}~\bibnamefont {Franzke}}, \bibinfo {author} {\bibfnamefont
  {H.}~\bibnamefont {Geissel}}, \bibinfo {author} {\bibfnamefont
  {T.}~\bibnamefont {Kerscher}}, \bibinfo {author} {\bibfnamefont
  {O.}~\bibnamefont {Klepper}}, \bibinfo {author} {\bibfnamefont {H.-J.}\
  \bibnamefont {Kluge}}, \bibinfo {author} {\bibfnamefont {C.}~\bibnamefont
  {Kozhuharov}}, \bibinfo {author} {\bibfnamefont {Y.}~\bibnamefont
  {Litvinov}}, \bibinfo {author} {\bibfnamefont {K.}~\bibnamefont {Löbner}},
  \bibinfo {author} {\bibfnamefont {G.}~\bibnamefont {Münzenberg}}, \bibinfo
  {author} {\bibfnamefont {N.}~\bibnamefont {Nankov}}, \bibinfo {author}
  {\bibfnamefont {F.}~\bibnamefont {Nolden}}, \bibinfo {author} {\bibfnamefont
  {Y.}~\bibnamefont {Novikov}}, \bibinfo {author} {\bibfnamefont
  {T.}~\bibnamefont {Ohtsubo}}, \bibinfo {author} {\bibfnamefont
  {T.}~\bibnamefont {Radon}}, \bibinfo {author} {\bibfnamefont
  {H.}~\bibnamefont {Schatz}}, \bibinfo {author} {\bibfnamefont
  {C.}~\bibnamefont {Scheidenberger}}, \bibinfo {author} {\bibfnamefont
  {M.}~\bibnamefont {Steck}}, \bibinfo {author} {\bibfnamefont
  {Z.}~\bibnamefont {Sun}}, \bibinfo {author} {\bibfnamefont {H.}~\bibnamefont
  {Weick}}, \ and\ \bibinfo {author} {\bibfnamefont {H.}~\bibnamefont
  {Wollnik}},\ }\href {\doibase https://doi.org/10.1023/A:1011911720453}
  {\bibfield  {journal} {\bibinfo  {journal} {Hyperfine Inter.}\ }\textbf
  {\bibinfo {volume} {132}},\ \bibinfo {pages} {289} (\bibinfo {year}
  {2001})}\BibitemShut {NoStop}%
\bibitem [{\citenamefont {Kimura}\ \emph {et~al.}(2018)\citenamefont {Kimura},
  \citenamefont {Ito}, \citenamefont {Kaji}, \citenamefont {Schury},
  \citenamefont {Wada}, \citenamefont {Haba}, \citenamefont {Hashimoto},
  \citenamefont {Hirayama}, \citenamefont {MacCormick}, \citenamefont
  {Miyatake}, \citenamefont {Moon}, \citenamefont {Morimoto}, \citenamefont
  {Mukai}, \citenamefont {Murray}, \citenamefont {Ozawa}, \citenamefont
  {Rosenbusch}, \citenamefont {Schatz}, \citenamefont {Takamine}, \citenamefont
  {Tanaka}, \citenamefont {Watanabe},\ and\ \citenamefont
  {Wollnik}}]{Kimura2018}%
  \BibitemOpen
  \bibfield  {author} {\bibinfo {author} {\bibfnamefont {S.}~\bibnamefont
  {Kimura}}, \bibinfo {author} {\bibfnamefont {Y.}~\bibnamefont {Ito}},
  \bibinfo {author} {\bibfnamefont {D.}~\bibnamefont {Kaji}}, \bibinfo {author}
  {\bibfnamefont {P.}~\bibnamefont {Schury}}, \bibinfo {author} {\bibfnamefont
  {M.}~\bibnamefont {Wada}}, \bibinfo {author} {\bibfnamefont {H.}~\bibnamefont
  {Haba}}, \bibinfo {author} {\bibfnamefont {T.}~\bibnamefont {Hashimoto}},
  \bibinfo {author} {\bibfnamefont {Y.}~\bibnamefont {Hirayama}}, \bibinfo
  {author} {\bibfnamefont {M.}~\bibnamefont {MacCormick}}, \bibinfo {author}
  {\bibfnamefont {H.}~\bibnamefont {Miyatake}}, \bibinfo {author}
  {\bibfnamefont {J.}~\bibnamefont {Moon}}, \bibinfo {author} {\bibfnamefont
  {K.}~\bibnamefont {Morimoto}}, \bibinfo {author} {\bibfnamefont
  {M.}~\bibnamefont {Mukai}}, \bibinfo {author} {\bibfnamefont
  {I.}~\bibnamefont {Murray}}, \bibinfo {author} {\bibfnamefont
  {A.}~\bibnamefont {Ozawa}}, \bibinfo {author} {\bibfnamefont
  {M.}~\bibnamefont {Rosenbusch}}, \bibinfo {author} {\bibfnamefont
  {H.}~\bibnamefont {Schatz}}, \bibinfo {author} {\bibfnamefont
  {A.}~\bibnamefont {Takamine}}, \bibinfo {author} {\bibfnamefont
  {T.}~\bibnamefont {Tanaka}}, \bibinfo {author} {\bibfnamefont
  {Y.}~\bibnamefont {Watanabe}}, \ and\ \bibinfo {author} {\bibfnamefont
  {H.}~\bibnamefont {Wollnik}},\ }\href@noop {} {\bibfield  {journal} {\bibinfo
   {journal} {Int. J. of Mass Spectrom.}\ }\textbf {\bibinfo {volume} {430}},\
  \bibinfo {pages} {134} (\bibinfo {year} {2018})}\BibitemShut {NoStop}%
\bibitem [{\citenamefont {Audi}\ \emph {et~al.}(2017)\citenamefont {Audi},
  \citenamefont {Kondev}, \citenamefont {Wang}, \citenamefont {Huang},\ and\
  \citenamefont {Naimi}}]{Audi2016}%
  \BibitemOpen
  \bibfield  {author} {\bibinfo {author} {\bibfnamefont {G.}~\bibnamefont
  {Audi}}, \bibinfo {author} {\bibfnamefont {F.~G.}\ \bibnamefont {Kondev}},
  \bibinfo {author} {\bibfnamefont {M.}~\bibnamefont {Wang}}, \bibinfo {author}
  {\bibfnamefont {W.~J.}\ \bibnamefont {Huang}}, \ and\ \bibinfo {author}
  {\bibfnamefont {S.}~\bibnamefont {Naimi}},\ }\href@noop {} {\bibfield
  {journal} {\bibinfo  {journal} {Chinese Phys. C}\ }\textbf {\bibinfo {volume}
  {41}},\ \bibinfo {pages} {030001} (\bibinfo {year} {2017})}\BibitemShut
  {NoStop}%
\bibitem [{\citenamefont {Bonatsos}\ \emph {et~al.}(2013)\citenamefont
  {Bonatsos}, \citenamefont {Karampagia}, \citenamefont {Cakirli},
  \citenamefont {Casten}, \citenamefont {Blaum},\ and\ \citenamefont
  {Susam}}]{Bonatsos2013}%
  \BibitemOpen
  \bibfield  {author} {\bibinfo {author} {\bibfnamefont {D.}~\bibnamefont
  {Bonatsos}}, \bibinfo {author} {\bibfnamefont {S.}~\bibnamefont
  {Karampagia}}, \bibinfo {author} {\bibfnamefont {R.~B.}\ \bibnamefont
  {Cakirli}}, \bibinfo {author} {\bibfnamefont {R.~F.}\ \bibnamefont {Casten}},
  \bibinfo {author} {\bibfnamefont {K.}~\bibnamefont {Blaum}}, \ and\ \bibinfo
  {author} {\bibfnamefont {L.~A.}\ \bibnamefont {Susam}},\ }\href {\doibase
  10.1103/PhysRevC.88.054309} {\bibfield  {journal} {\bibinfo  {journal} {Phys.
  Rev. C}\ }\textbf {\bibinfo {volume} {88}},\ \bibinfo {pages} {054309}
  (\bibinfo {year} {2013})}\BibitemShut {NoStop}%
\bibitem [{\citenamefont {Van~Isacker}\ \emph {et~al.}(1999)\citenamefont
  {Van~Isacker}, \citenamefont {Juillet},\ and\ \citenamefont
  {Nowacki}}]{VanIsacker1999}%
  \BibitemOpen
  \bibfield  {author} {\bibinfo {author} {\bibfnamefont {P.}~\bibnamefont
  {Van~Isacker}}, \bibinfo {author} {\bibfnamefont {O.}~\bibnamefont
  {Juillet}}, \ and\ \bibinfo {author} {\bibfnamefont {F.}~\bibnamefont
  {Nowacki}},\ }\href {\doibase 10.1103/PhysRevLett.82.2060} {\bibfield
  {journal} {\bibinfo  {journal} {Phys. Rev. Lett.}\ }\textbf {\bibinfo
  {volume} {82}},\ \bibinfo {pages} {2060} (\bibinfo {year}
  {1999})}\BibitemShut {NoStop}%
\bibitem [{\citenamefont {Hasegawa}\ \emph {et~al.}(2007)\citenamefont
  {Hasegawa}, \citenamefont {Kaneko}, \citenamefont {Mizusaki},\ and\
  \citenamefont {Sun}}]{Hasegawa2007}%
  \BibitemOpen
  \bibfield  {author} {\bibinfo {author} {\bibfnamefont {M.}~\bibnamefont
  {Hasegawa}}, \bibinfo {author} {\bibfnamefont {K.}~\bibnamefont {Kaneko}},
  \bibinfo {author} {\bibfnamefont {T.}~\bibnamefont {Mizusaki}}, \ and\
  \bibinfo {author} {\bibfnamefont {Y.}~\bibnamefont {Sun}},\ }\href {\doibase
  https://doi.org/10.1016/j.physletb.2007.09.017} {\bibfield  {journal}
  {\bibinfo  {journal} {Phys. Lett. B}\ }\textbf {\bibinfo {volume} {656}},\
  \bibinfo {pages} {51 } (\bibinfo {year} {2007})}\BibitemShut {NoStop}%
\end{thebibliography}%
 






\end{document}